\documentclass[10pt,oneside,final]{IEEEtran}

\usepackage{amsfonts}
\usepackage{amsmath}
\usepackage{graphicx}
\usepackage{color}
\usepackage{multicol}
\usepackage{amssymb}
\usepackage{subfigure}
\usepackage{bm}
\usepackage{latexsym}
\usepackage{stfloats}
\usepackage{float}
\usepackage{exscale}
\usepackage{relsize}
\usepackage{cite}
\usepackage{mathtools}

\usepackage{algorithm}
\usepackage{algorithmicx}
\usepackage{algpseudocode}

\usepackage{setspace}
\usepackage{url}

\begin{document}

\title{Deep Reinforcement Learning for RIS-aided Multiuser Full-Duplex Secure Communications with Hardware Impairments}
\IEEEoverridecommandlockouts

\author{Zhangjie~Peng,
        Zhibo~Zhang,
        Lei~Kong,\\
        Cunhua Pan,~\IEEEmembership{Member,~IEEE,}
        Li~Li,
        and Jiangzhou~Wang,~\IEEEmembership{Fellow,~IEEE}
\thanks{
	This work was supported in part by the Natural Science Foundation of Shanghai under Grant 22ZR1445600, in part by the open research fund of National Mobile Communications Research Laboratory, Southeast University under Grant 2018D14, and in part by the National Natural Science Foundation of China under Grant 61701307. \emph{ (Corresponding authors: Cunhua Pan and Zhibo Zhang.)}
				
	Zhangjie Peng is with the College of Information, Mechanical and Electrical Engineering, Shanghai Normal University, Shanghai 200234, China, also with the National Mobile Communications Research Laboratory, Southeast University, Nanjing 210096, China, and also with the Shanghai Engineering Research Center of Intelligent Education and Bigdata, Shanghai Normal University, Shanghai 200234, China (e-mail: pengzhangjie@shnu.edu.cn).
	
	Zhibo Zhang and Li Li are with the College of Information, Mechanical and Electrical Engineering, Shanghai Normal University, Shanghai 200234, China (e-mail: 1000497171@smail.shnu.edu.cn; lilyxuan@shnu.edu.cn). 
	
	Lei Kong is with New H3C Technologies Co., Limited, Hangzhou 310052, China, and also with the National Mobile Communications Research Laboratory, Southeast University, Nanjing 210096, China (e-mail: konglei.seu@gmail.com).
	
	Cunhua Pan is with the National Mobile Communications Research Laboratory,
	Southeast University, Nanjing 210096, China (e-mail: cpan@seu.edu.cn). 
	
	Jiangzhou Wang is with the School of Engineering, University of Kent, Canterbury CT2 7NT, U.K. (e-mail:
	j.z.wang@kent.ac.uk).
	
	Copyright (c) 20xx IEEE. Personal use of this material is permitted. However, permission to use this material for any other purposes must be obtained from the IEEE by sending a request to pubs-permissions@ieee.org.
}
}
%
\maketitle
\newtheorem{lemma}{Lemma}
\newtheorem{theorem}{Theorem}
\newtheorem{remark}{Remark}
\newtheorem{corollary}{Corollary}
\newtheorem{proposition}{Proposition}
\begin{abstract}
In this paper, we investigate a reconfigurable intelligent surface (RIS)-aided multiuser full-duplex secure communication system with hardware impairments at transceivers and RIS, where multiple eavesdroppers overhear the two-way transmitted signals simultaneously, and an RIS is applied to enhance the secrecy performance. Aiming at maximizing the sum secrecy rate (SSR), a joint optimization problem of the transmit beamforming at the base station (BS) and the reflecting beamforming at the RIS is formulated under the transmit power constraint of the BS and the unit modulus constraint of the phase shifters. As the environment is time-varying and the system is high-dimensional, this non-convex optimization problem is  mathematically intractable. A deep reinforcement learning (DRL)-based algorithm is explored to obtain the satisfactory solution by repeatedly interacting with and learning from the dynamic environment. Extensive simulation results illustrate that the DRL-based secure beamforming algorithm is proved to be significantly effective in improving the SSR. It is also found that the performance of the DRL-based method can be greatly improved and the convergence speed of neural network can be accelerated with appropriate neural network parameters.
\end{abstract}

\begin{IEEEkeywords}
Reconfigurable intelligent surface (RIS), secure communication, full-duplex, hardware impairment, deep reinforcement learning (DRL).

\end{IEEEkeywords}

\vspace{-0.4cm}
\section{Introduction}
\IEEEPARstart{R}{econfigurable} intelligent surface (RIS), also known as intelligent reflecting surface (IRS), is one of the most promising and disruptive technologies in the sixth generation (6G) and beyond wireless communication
\cite{8796365,8910627,9475160,zhou2020service}. 
The RIS is a uniform planar array made of numerous low-cost and nearly-passive components. Every unit can reflect the incident electromagnetic wave independently to the desired direction by dynamically adjusting the phase shifts or amplitudes with the controller \cite{9211520}. 
RIS can be expediently installed on ceilings or billboards to establish a virtual line-of-sight (LoS) link between the source and the destination when the direct link is blocked by obstacles \cite{9475160,9090356}.

Due to the inherent broadcast nature of wireless communication, the transmitted information is easy to be eavesdropped. Thus, physical layer security (PLS) is an indispensable part of the wireless communication, which attracts a lot of attention. 
Recently, RIS 
has been introduced into secure communication as an effective way with low power consumption and anti-eavesdropping. With the assistance of RIS, the reflected signals received by the legitimate users will be constructively enhanced while the signals leaked to the eavesdroppers will be destructively weakened. Extensive research attention has been devoted to RIS-assisted PLS. The initial contributions were based on the simple system model with only one legitimate user in the presence of an eavesdropper 
\cite{8723525,8743496,8847342}. 
While in \cite{9293148} and \cite{9291402}, a legitimate user was wiretapped by multiple eavesdroppers. 
Different from these contributions considering a single legitimate user \cite{8723525,8743496,8847342,9293148,9291402}, an RIS-aided secure system including multiple legitimate users and eavesdroppers was considered in \cite{9133130,8742603}. 

Most of the existing literatures on RIS-assisted PLS \cite{8723525,8743496,8847342,9293148,9291402,9133130,8742603} considered half-duplex (HD) communications, in which the signals cannot be transmitted and received at the same time and over the same frequency. However, it is less difficult for an eavesdropper to decode the desired signal from the eavesdropped signal in an HD communication system \cite{8644299}. The development of full-duplex (FD) technology in wireless communication provides the possibility of enhancing PLS \cite{8644299}. Meanwhile, FD communications enable the transceivers to exchange information simultaneously in the same frequency band, which can effectively improve spectrum efficiency with the development of residual self-interference cancellation  technology. Recently, RIS-assisted FD systems were studied in \cite{9158342,9318531}, which highlighted the the superiority
of RIS-aided FD communications under proper residual self-interference cancellation. Furthermore, in RIS-aided FD secure communication systems, the bidirectional signals can be simultaneously received and transmitted in the same frequency band, which can theoretically lead to the overlap of multiple signals at the eavesdroppers to improve the secrecy performance \cite{9248012}. Specifically, when the uplink/downlink signals are eavesdropped, the downlink/uplink signals reflected by the RIS will be reused as interference noise to degrade the eavesdropping performance of the eavesdroppers. To our knowledge, the potential of deploying RIS to secure FD communications has rarely been explored \cite{9248012,9383283}. The closed-form expression for the sum secrecy rate (SSR) of an RIS-aided multi-pair two-way  communication system was derived in \cite{9248012}. In \cite{9383283}, the SSR was maximized by optimizing the transmit power and the phase shifts. However, there is still lack of research work on RIS-assisted multiuser FD secure communications. 

It is worth noting that the above contributions on RIS-assisted secure communications \cite{8723525,8743496,8847342,9293148,9291402,9133130,8742603,9383283,9248012} were based on the assumption that the hardware is perfect. 
However, 
transceiver hardware impairment (T-HWI) is generally inevitable 
considering the non-linearities and phase noise, $et\ al.$ \cite{6891254,9298472}. Even if the mitigation/compensation algorithm was proposed, T-HWI cannot be completely eliminated \cite{6891254}. A robust  beamforming designed algorithm was proposed in \cite{9239335} to maximize the received signal-to-noise ratio (SNR) considering the impact of T-HWI. 
The robust transmission optimization of an RIS-assisted secure communication system with T-HWI was studied for the first time in \cite{9374557}. At the same time, the high-precision configuration of the phase shifts at the RIS is not feasible \cite{8869792}, thus there is also the HWI at the RIS (RIS-HWI). 
The achievable rate expression of RIS-assisted mmWave communication system was derived, taking into account the RIS-HWI \cite{9295369}. 
The impact of RIS-HWI on the performance of a secure communication system was investigated in \cite{9219155}.
However, there were only a few studies that analysed the RIS-aided communication systems with both the T-HWI and the RIS-HWI. Both analytical and experimental results in \cite{9390410} showed that RIS-HWI as well as T-HWI reduced the achievable rate. 
The spectral and energy efficiency was analyzed in \cite{9079457} considering an RIS-aided MISO communication system with both T-HWI and RIS-HWI.

All the above-mentioned contributions \cite{8723525,8743496,8847342,9293148,9291402,9133130,8742603,9158342,9318531,9383283,9248012,6891254,9298472,9239335,9374557,8869792,9295369,9219155,9390410,8644299,9079457} were based on traditional mathematical optimization methods such as alternating optimization (AO) and block coordinate descent (BCD) method to design the transmit beamforming and/or phase shifts, which are model-based design paradigms that require accurate mathematical models and expert knowledge. However, traditional model-based wireless techniques are hard to meet the demanding requirements of emerging applications in future 6G networks, including excessively complex communication scenarios with unknown channel model and communication scenarios that cannot be described mathematically such as the non-linearity due to the inevitable HWI \cite{9165550}. On the contrary, deep learning (DL) is a powerful technique capable of learning complex interrelationships among variables, especially those that are difficult to accurately describe with mathematical models, which empowers us to design wireless communication systems without knowledge of precise mathematical models \cite{9165550}. In addition, dynamic 6G networks can lead to uncertainty in the wireless environment (for example, due to the changing of RIS configuration), which brings great difficulties to real-time sensing.  Meanwhile, artificial intelligence (AI) techniques are more robust against the uncertainty in the systems, which can achieve accurate real-time sensing. Therefore, the application of AI to 6G networks can optimize the network structure and improve the communication performance \cite{9237460}. Inspired by the advantages of model-free AI, extensive research attention has been devoted to apply DL to the phase shifts design \cite{8815412,9370097}.

However, DL methods that have been applied in the existing contributions \cite{8815412,9370097} to solve the sophisticated problems are supervised learning. 
Although supervised learning has shown promising results in wireless problems such as channel estimation, RIS phase shifts design and so on, it requires a large number of prior labeled training and test dataset for offline training, which relies on either the AO algorithms or the exhaustive search. As known, the training dataset is difficult to obtain in many circumstances.
Furthermore, for an RIS-aided communication system, once the number of phase shifters changes, numerous training samples need to be reacquired and the DL network need to be retrained from scratch, which greatly limits the application of DL-based method.
In the considered sophisticated optimization problem, the artificially labeled training dataset is unable to be acquired. On the contrary, as a branch of machine learning (ML), reinforcement learning (RL) has been considered to be an effective approach to deal with complex control tasks, which possesses the characteristics of online learning and sample generation, and does not require the training labels with prior knowledge. Furthermore, deep reinforcement learning (DRL), which combines the advantages of DL in function fitting and RL in policy-making, has been leveraged to optimize the phase shifts \cite{8968350,9110869,9410457}. Recently, DRL-based methods have also been applied to solve the optimization problem for RIS-aided secure communications \cite{9206080,9264659}. 
On the other hand, due to the time-varying channels and the high-dimensional system, the proposed non-convex optimization problem is mathematically intractable. It is challenging to derive the explicit mathematical 
optimal solution for this problem. Fortunately, no complicated mathematical formulations is necessary for DRL, which can significantly reduce the computational complexity and computing time.
Meanwhile, 
DRL is able to learn the implicit knowledge about the radio channels through the trial-and-error interactions without knowing the channel model and mobility pattern. Therefore, DRL is particularly beneficial to deal with the problem in a high-dimensional wireless communication system where the radio channel varies over time. 
Motivated by the aforementioned facts, we explore a DRL-based method to address the proposed sophisticated optimization problem.

In this paper, we present an RIS-assisted FD two-way wiretap communication system with multiple legitimate users as well as multiple eavesdroppers. Both T-HWI and RIS-HWI are considered. 
The transmit beamforming at the BS and phase shifts at the RIS are jointly optimized to maximize the SSR. 
This complex optimization problem is non-convex and the optimal solution is unknown.
A novel DRL-based method is developed to obtain the satisfactory solution, in contrast to numerical optimization algorithms by utilizing complicated mathematical formulations. The main contributions of this paper are summarized as follows:

\begin{itemize}
  \item  To the best of our knowledge, this is the first attempt to study the performance of an RIS-assisted FD secure communication system with both T-HWI and RIS-HWI. In order to improve the secrecy performance, an optimization problem for jointly designing the transmit and reflecting beamforming is formulated. In the complex and time-varying environment, it is a mathematically intractable non-convex optimization problem, which is modeled as a Markov decision process (MDP). 
  \item  A DRL-based secure beamforming algorithm is developed to jointly design the transmit and reflecting beamforming to maximize the SSR. Specifically, the SSR is used as the instant reward to train the DRL network. The training data is generated online through the trial-and-error interactions between the agent and the environment. The network parameters are adjusted accordingly, so as to design the transmit and reflecting beamforming simultaneously by gradually maximizing the SSR through repeated iterations. Since the action space containing the transmit beamforming matrix and the phase shifts matrix is continuous, DDPG technique is adopted.
  \item  Dispensing with explicit mathematical calculation, the proposed DRL-based algorithm has a standard framework, as well as low complexity in its implementation,  which can be conveniently deployed to the communication system with different settings. On the other hand, compared with the DL-based algorithm that requires artificially labeled training datasets, the DRL-based algorithm learns the knowledge of the environment to adapt.
  \item Extensive simulation results illustrate that the proposed DRL-based algorithm is significantly effective in improving the SSR. Specifically, the agent improves its action policy step by step according to the reward fed back from the environment, to obtain the satisfactory transmit beamforming and phase shifts to improve the SSR. It is also found that the performance of the DRL-based method can be greatly improved and the convergence speed of neural network can be accelerated with appropriate neural network parameters.
\end{itemize}

The remainder of this paper is organized as follows. In Section \ref{section 2}, we introduce the RIS-aided FD secure communication system with HWI and formulate the SSR maximization problem. In Section \ref{section 4}, we present the proposed DRL-based algorithm for joint optimization of the transmit beamforming and the phase shifts. In Section \ref{section 5}, we provide extensive simulation results to elaborate the performance of the developed DRL-based algorithm. In Section \ref{section 6}, we draw a brief conclusion.

\section{System Model and Problem Formulation}
\label{section 2}
\subsection{System Model}

\begin{figure}[t]
	\centering
	\includegraphics[scale=0.52]{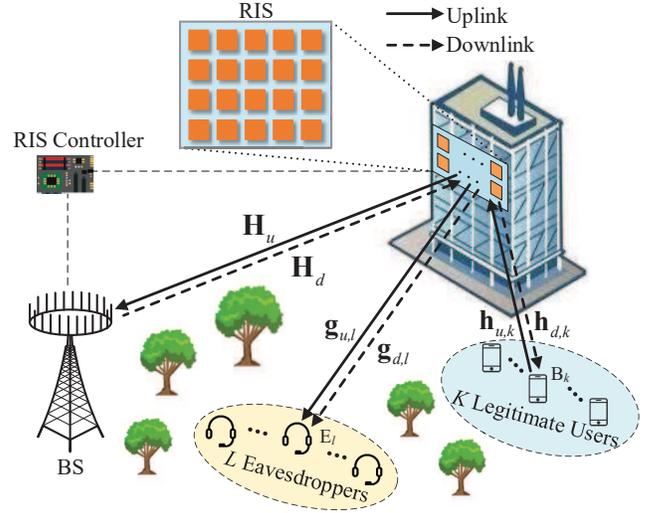}
	\caption{System model for RIS-aided multiuser FD two-way secure communication system with multi-eavesdropper.}
	\label{model}
\end{figure}

Consider an RIS-assisted FD two-way wiretap communication system, as illustrated in Fig. \ref{model}. Both the BS and the legitimate users work in the FD mode.
The BS transmits independent confidential information to $K$ legitimate users, while each legitimate user sends a data stream to the BS. At the same time, $L$ unauthorized eavesdroppers that are distributed around the BS and legitimate users are trying to wiretap the transmission information from the BS and the legitimate users independently. An RIS is applied to enhance the secrecy performance. Moreover, due to the existence of many obstacles, the signals are only transmitted via the RIS-assisted reflecting links.
We denote the $k$-th legitimate user as $\text B_{k}$ and the $\ell$-th eavesdropper as $\text E_{\ell}$, for $k=1, \ldots, K$ and $\ell=1, \ldots, L$. Each legitimate user has a single transmit antenna and a single receive antenna, while each eavesdropper is equipped with one antenna \cite{7934326}. 
The BS is equipped with $N_t$ transmit antennas and $N_r$ receive antennas. The RIS consists of $M$ passive reflecting elements. The phase shift of each reflecting element can be dynamically adjusted by a smart controller coordinated with the BS. The phase shifts matrix $\bm{\Theta}$ is denoted by $\bm{\Theta} = {\rm{diag}}({\chi _1}{e^{j{\theta _1}}}, \ldots, {\chi _m}{e^{j{\theta _m}}}, \ldots, {\chi _M}{e^{j{\theta _M}}})$, where ${\chi _m} \in \left[ {0,1 } \right]$ and ${\theta _m} \in \left[ {0,2\pi } \right]$ represent the amplitude and phase shift of the $m$-th reflecting element, respectively.

Channel state information (CSI) is crucial for achieving the performance gain of the RIS-assisted secure communication. We suppose that the CSI of all channels is perfectly known at the BS before the data transmission.  On the one hand, the acquisition of CSI for the legitimate users has been widely explored \cite{9366805,8937491, 9400843}. 
On the other hand, for the CSI of the eavesdroppers, if the eavesdroppers are active in the secure transmission system\footnote{For example, the eavesdroppers can be disguised as active users in the secure communication system trusted by the BS but untrusted by the legitimate users.}, then their CSI acquisition can be acquired by regarding them as the legitimate users. If the eavesdroppers are passive\footnote{The eavesdroppers are passive, which means that they never transmit.},
their CSI can also be obtained by some technologies, such as detecting the local oscillator power accidentally leakage from the eavesdropper receivers’ RF front end \cite{6288501,9262884}.

According to \cite{8811733}, we assume that the power of the signal reflected twice or more times by the RIS is imperceptible, and therefore is ignored. In addition, RIS is designed to maximize the power of the incident signal so that there is no energy loss during reflection.


\subsection{Channel Model}
According to the above-mentioned eavesdroppers' CSI acquisition method, we suppose that the BS perfectly knows the eavesdroppers' CSI. The baseband equivalent channels BS $ \to $ RIS, RIS $ \to $ BS, RIS $ \to $ $k$-th legitimate user $\text B_{k}$, $\text B_{k}$ $ \to $ RIS , the downlink RIS $ \to $ $\ell$-th eavesdropper $\text E_{\ell}$, and the uplink RIS $ \to $ $\text E_{\ell}$ are represented as ${{\bf{H}}_{d}} \in {{\mathbb C}^{M \times N_t}}$, ${{\bf{H}}_{u}} \in {{\mathbb C}^{M \times N_r}}$, ${{\bf{h}}_{d,k}} \in {{\mathbb C}^{M \times 1}}$, ${{\bf{h}}_{u,k}} \in {{\mathbb C}^{M \times 1}}$, ${{\bf{g}}_{d,\ell}} \in {{\mathbb C}^{M \times 1}}$, ${{\bf{g}}_{u,\ell}} \in {{\mathbb C}^{M \times 1}}$, respectively, as shown in Fig. \ref{model}.

It is assumed that the RIS is installed on the outer wall of a high building with only a few scatters. Therefore, 
we adopt a Rician fading channel model\footnote{Actually, the proposed algorithm is suitable for any channel model.} for all reflecting channels, which is composed of LoS components as well as NLoS components. The channels can be expressed as
\begin{align}\label{1}
{\bf{h}} = \sqrt {\alpha_{\bf{h}}}  \left( {\sqrt {\frac{\varepsilon_{\bf{h}} }{{\varepsilon_{\bf{h}}  + 1}}} \overline {\bf{h}}  + \sqrt {\frac{1}{{\varepsilon_{\bf{h}}  + 1}}} \widetilde {\bf{h}}} \right),
\end{align}
where ${\bf{h}} \in {\bf{\mathcal {H}}} = \left\{ {{{\bf{H}}_{d}},{{\bf{H}}_u},{{\bf{h}}_{d,k}},{{\bf{h}}_{u,k}},{{\bf{g}}_{d,\ell}},{{\bf{g}}_{u,\ell}}} \right\}$, $\alpha_{\bf{h}}$ denotes the large-scale fading coefficient, $\varepsilon_{\bf{h}}$ denotes the Rician factor. ${\overline{\bf{h}}}$ is deterministic LoS channel component. ${\widetilde{\bf{h}}}$ is the NLoS channel component whose elements are independent and identical distribution (i.i.d.) standard Gaussian random variables following the distribution of $\mathcal{CN}(0,1)$.

The BS and RIS are both equipped with uniform linear array (ULA), 
the LoS components 
are modeled as
\begin{align}\label{2}
\overline {\bf{h}}  = {{\bf{a}}_{{\bf{h}}, r}}\left( {{\vartheta ^{{\bf{h}}, {\rm{AOA}}}}} \right){\bf{a}}_{{\bf{h}}, t}^{\rm{H}}\left( {{\vartheta ^{{\bf{h}}, {\rm{AOD}}}}} \right),
\end{align}
where ${{\bf{a}}_{{\bf{h}}, r}}\left( {{\vartheta ^{{\bf{h}}, {\rm{AOA}}}}} \right)$ and ${\bf{a}}_{{\bf{h}}, t}^{\rm{H}}\left( {{\vartheta ^{{\bf{h}}, {\rm{AOD}}}}} \right)$ are the array responses of  $W_{\bf{h}}$-element/antenna ULA, which are respectively given by
\begin{align}\label{3}
\!\!\!{{\bf{a}}_{{\bf{h}}, r}}\left( {{\vartheta ^{{\bf{h}}, {\rm{AOA}}}}} \right) \!\!=\!\! \left[ {1, \ldots, {e^{j2\pi \frac{d_{\bf{h}}}{\lambda_{\bf{h}} }\left( {{W_{{\bf{h}}, r}} - 1} \right)\sin {\vartheta ^{{\bf{h}}, {\rm{AOA}}}}}}} \right]^T,
\end{align}
\begin{align}
\!\!\!\!{{\bf{a}}_{{\bf{h}}, t}}\left( {{\vartheta ^{{\bf{h}}, {\rm{AOD}}}}} \right) \!\!=\!\! \left[ {1, \ldots, {e^{j2\pi \frac{d_{\bf{h}}}{\lambda_{\bf{h}} }\left( {{W_{{\bf{h}}, t}} - 1} \right)\sin {\vartheta ^{{\bf{h}}, {\rm{AOD}}}}}}} \right]^T,
\end{align}
where $W_{{\bf{h}}, r}$ and $W_{{\bf{h}}, t}$ represent the numbers of antennas/elements at the receiver and transmitter, respectively\cite{9318531}, ${\vartheta ^{{\bf{h}}, {\rm{AOA}}}}$ and ${\vartheta ^{{\bf{h}}, {\rm{AOD}}}}$ are the angle of arrival
(AoA) and angle of departure (AoD), respectively, $d_{\bf{h}}$ is the inter-element spacing at the ULA, $\lambda_{\bf{h}}$ is the carrier wavelength.

\subsection{Hardware Impairment Model}

In practice, due to the non-ideality of hardware, the transmitted and received signals are affected by HWI, which generally exists in practical communication systems, including RIS-aided systems with RIS-HWI \cite{9390410}. In the considered system, there are two kinds of HWI with different mathematical models, including RIS-HWI and T-HWI.

Firstly, RIS-HWI is caused by the inherent hardware imperfections of the reflecting elements or the imperfect channel estimations, which is also called phase noise \cite{8869792}. The phase noise matrix is expressed as ${\bf{\Phi }} = {\rm{diag}}\left( {{e^{j\Delta {\theta _1}}}, \ldots, {e^{j\Delta {\theta _m}}}, \ldots, {e^{j\Delta {\theta _M}}}} \right)$, where $\Delta {\theta _m}$ is random phase noise uniformly distributed within $\left[ {{{ - \pi } \mathord{\left/
 {\vphantom {{ - \pi } {2,{\pi  \mathord{\left/
 {\vphantom {\pi  2}} \right.
 \kern-\nulldelimiterspace} 2}}}} \right.
 \kern-\nulldelimiterspace} {2,{\pi  \mathord{\left/
 {\vphantom {\pi  2}} \right.
 \kern-\nulldelimiterspace} 2}}}} \right]$ for $m=1,2, \ldots, M$, according to \cite{9390410}.

Then, another kind of HWI, namely T-HWI, is modeled as an independent Gaussian distortion noise, which will cause that the received signal dose not match the expected signal or produce distortion on the received signal during signal processing \cite{9390410}. The distortion power of the transmit antenna increases with the increase of the signal power allocated to the antenna \cite{6891254}. 
\begin{figure*}[b]
	\setcounter{equation}{10}
	\hrulefill
	\begin{align}\label{11}
		\Gamma_{k}^{\rm B} \left( {{\bf{W}},{\bf{\Theta }}} \right) =&  \sum\limits_{i = 1,i \ne k}^K {{\left| {{\bf{h}}_{d,k}^{\rm{H}}{\bf{\Theta \Phi }}{{\bf{H}}_d}{{\bf{w}}_i}} \right|}^2} + \kappa _d^{\rm{S}}{\bf{h}}_{d,k}^{\rm{H}}{\bf{\Theta \Phi }}{{\bf{H}}_{{d}}}\widetilde {{\rm{diag}}}\left\{ {\sum\limits_{i = 1}^K {{{\bf{w}}_i}} {\bf{w}}_{{i}}^{\rm{H}}} \right\}{\bf{H}}_{{d}}^{\rm{H}}{{\bf{\Phi }}^{\rm{H}}}{{\bf{\Theta }}^{\rm{H}}}{{\bf{h}}_{d,k}} + \sigma _{d,k}^2 \nonumber\\ &+ \sum\limits_{i = 1}^K {\left( {1 + \kappa _{u,i}^{\rm{B}}} \right)\rho {P_i}{{\left| {{\bf{h}}_{_{u,k}}^{\rm{H}}{\bf{\Theta \Phi }}{{\bf{h}}_{u,i}}} \right|}^2} }    .
	\end{align}
\end{figure*}

\subsection{Signal Transmission Model}
At the BS side, let ${\bf{w}}_k\in {{\mathbb C}^{N_t \times 1}}$ be the beamforming vector for $\text B_{k}$ and ${{s}}_{d,k}$ be the confidential information sent to $\text B_{k}$. Therefore, the signal transmitted from the BS is modeled as
\begin{align}\label{5}
	\setcounter{equation}{5}
{{\bf{x}}_d} =&~ {\widehat {\bf{x}}_d} + {\bm{\eta}}_d^{\rm{S}}, \tag{5a}
\end{align}
\begin{align}
{\widehat {\bf{x}}_d} =&~ \sum\limits_{k = 1}^K {{{\bf{w}}_k}{s_{d,k}}},\tag{5b}
\end{align}
where ${{s}}_{d,k}$ is modeled as i.i.d. random variable with zero mean and unit variance, i.e. ${{s}}_{d,k}\sim\mathcal{CN}(0,1)$ and $\mathbb{E}\{{\left| s_{d,k} \right|^2}\}=1$. According to 
\cite{9394419},
${\bm{\eta}}_d^{\rm{S}}\in {{\mathbb C}^{N_t \times 1}}$ is the independent Gaussian transmit distortion noise. The power of the distortion noise at each antenna is proportional to the transmit power at that antenna, which can be modeled as $\bm{\eta} _d^{\rm{S}} \sim \mathcal{CN} \left( {\bm{0},\bm{\Upsilon} _d^{\rm{S}}} \right)$, where \cite{9239335}
\begin{align}\label{7}
\bm{\Upsilon} _d^{\rm{S}} = 
\kappa _d^{\rm{S}}\widetilde {\rm{diag}} \left\{\sum\limits_{k = 1}^{\rm{K}} {{ {{{\bf{w}}_k}}{{{\bf{w}}_{k}^{\rm H}}}}}\right\},
\end{align}
with $\kappa _d^{\rm{S}} \ge 0$ being a scale factor that describes the severity of HWI at the BS's transmitter, $\widetilde{\rm{diag}}{\{{\bf X}\}}$ denotes a diagonal matrix whose diagonal entries are the diagonal elements of matrix $\bf{X}$. The transmit power at the BS is subject to the maximum power constraint
\begin{align}\label{8}
{\rm{Tr}}\left( {{\bf{W}}{{\bf{W}}^{\rm H}}} \right) \le {P_{max }},
\end{align}
where ${\bf{W}} \buildrel \Delta \over = \left[ {{{\bf{w}}_1},{{\bf{w}}_2}, \ldots, {{\bf{w}}_K}} \right] \in {{\mathbb C}^{N_t \times K}}$, and $P_{max }$ is maximum transmit power of the BS.

Similarly, at the legitimate users side, the transmitted signal from $B_k$ is
\begin{align}\label{9}
	\setcounter{equation}{8}
{x_{u,k}} =&~ \sqrt {{P_k}} {{{s}}_{u,k}},\tag{8a}
\end{align}
\begin{align}
{{{s}}_{u,k}} =&~ {\widehat {{s}}_{u,k}} + \eta _{u,k}^{\rm{B}},\tag{8b}
\end{align}
where ${\widehat{{{s}}}_{u,k}}\sim \mathcal{CN}(0,1)$ represents the independent Gaussian uplink information symbol sent by $B_k$, ${P_k}$ is the transmit power of $B_k$, $\eta _{u,k}^{\rm{B}}$ is an independent Gaussian transmit distortion noise and $\eta _{u,k}^{\rm{B}}\sim \mathcal{CN}(0,\kappa _{u,k}^{\rm{B}}\mathbb{E}\left\{ {{{\widehat {{s}}}_{u,k}}\widehat {{s}}_{u,k}^{\rm{H}}} \right\})$ , where $\kappa _{u,k}^{\rm{B}} \ge 0$ represents a scale factor that describes the severity of HWI at the $k$-th legitimate user's transmitter.

For downlink, the received signal at $B_k$ is written as
\begin{align}\label{YB}
y_k^{\rm{B}} =& \underbrace {{\bf{h}}_{d,k}^{\rm{H}}{\bf{\Theta \Phi }}{{\bf{H}}_d} {{{\bf{w}}_k}{{{s}}_{d,k}}} }_{\text{Desired\ signal}} +\!\!\! \underbrace {\sum\limits_{i = 1,i \ne k}^K {{\bf{h}}_{_{d,k}}^{\rm{H}}{\bf{\Theta \Phi }}{{\bf{H}}_d} {{{\bf{w}}_i}{{{s}}_{d,i}} } } }_{\text{Multiuser\ interference}}
\nonumber\\
 & + \underbrace {{\bf{h}}_{d,k}^{\rm{H}}{\bf{\Theta \Phi }}{{\bf{H}}_d} {{\bm{\eta }}_d^{\rm{S}}} }_{\text{Transmitter\ HWI}}+ \underbrace {\sqrt {{\rho _L}} \sqrt {{P_k}} {h_{kk}}\left( {{{\widehat {{s}}}_{u,k}} + \eta _{u,k}^{\rm{B}}} \right)}_{\text{Loop-interference}}\nonumber\\ &+ \underbrace {\sqrt {{\rho _S}} \sqrt {{P_k}} {\bf{h}}_{_{d,k}}^{\rm{H}}{\bf{\Theta \Phi }}{{\bf{h}}_{u,k}}\left( {{{\widehat {{s}}}_{u,k}} +\! \eta _{u,k}^{\rm{B}}} \right)}_{\text{Self-interference}}~+\!\!\!\!\!\underbrace{\eta _{d,k}^{\rm{B}}}_{\text{Receiver HWI}} \nonumber\\ &+\!\!\!\! \underbrace {\sum\limits_{i = 1,i \ne k}^K {\sqrt {{P_i}} {\bf{h}}_{_{d,k}}^{\rm{H}}{\bf{\Theta \Phi }}{{\bf{h}}_{u,i}}\left( {{{\widehat {{s}}}_{u,i}} + \eta _{u,i}^{\rm{B}}} \right)} }_{\text{Co-channel\ interference}}+ \underbrace {{{n}}_k^{\rm{B}}}_{\text{Noise}},
\end{align}
where $h_{kk}$ is the loop channel between the transmit antenna and receive antenna of the $k$-th legitimate user $B_k$, ${{{n}}_k^{\rm{B}}}$ is additive white Gaussian noise (AWGN) following the distribution of $\mathcal{CN}(0,\sigma _{k}^2)$, and ${\eta _{d,k}^{\rm{B}}}$ is an independent additional distortion noise term at the $k$-th legitimate user's receiver. 
Similar to \cite{9374557}, we denote the received signal as $y_k^{\rm{B}} = \widetilde {y}_k^{\rm{B}} + \eta _{d,k}^{\rm{B}}$, then ${\eta _{d,k}^{\rm{B}}}$ follows the distribution of $\mathcal{CN}(0,\kappa _{d,k}^{\rm{B}}\mathbb{E}\left\{ {{\left| \widetilde {y}_k^{\rm{B}} \right|}^2} \right\})$, where $\kappa_{d,k}^{\rm B}$ is a proportionality coefficient that describes the severity of HWI at the $k$-th legitimate user's receiver. The coefficients $\rho _L$ and $\rho _S$ with $0 \le {\rho _L},{\rho _S} \le 1$ are the coefficients of loop-interference (LI) and self-interference (SI), respectively. $\rho_L$ and $\rho_S$ are introduced to characterize the LI and SI which cannot be completely eliminated by some interference elimination techniques, respectively. 
According to \cite{9318531}, LI term and noise term can be combined, the sum of which is denoted as ${{m}}_{d,k}$, whose average
power is written as $\sigma _{d,k}^2 = {\left| {{m_{d,k}}} \right|^2} = {\rho _L}{P_k}{\left| {{h_{kk}}} \right|^2} + \sigma _k^2$. Accordingly, the signal-to-interference-plus-noise ratio (SINR) at $B_k$ is
\begin{align}\label{SINRB}
{\rm{SINR}}_k^{\rm{B}} = \frac{{{{\left| {{\bf{h}}_{d,k}^{\rm{H}}{\bf{\Theta \Phi }}{{\bf{H}}_d}{{\bf{w}}_k}} \right|}^2}}}{{\Gamma_k^{\rm B} \left( {{\bf{W}},{\bf{\Theta }}} \right) + {{\left| {\eta _{d,k}^{\rm{B}}} \right|}^2}}},
\end{align}
where $\Gamma_k^{\rm B} \left( {{\bf{W}},{\bf{\Theta }}} \right)$ is given by \eqref{11} at the bottom of this page,
where the coefficient $\rho$ is defined as
$\rho  = \left\{ {\begin{array}{*{20}{c}}
{{\rho _{\rm{S}}},\ \ {\text{if }}\ i = k;}\\
{1,\ \ {\text{otherwise}}{.}}
\end{array}} \right.$
\begin{figure*}[b]
	\setcounter{equation}{14}
	\hrulefill
	\begin{align}\label{15}
		\hspace{-0.3cm}{\Gamma _{k}^{\rm S}}\left( {{\bf{W}},{\bf{\Theta }}} \right) =  {\sum\limits_{i = 1,i \ne k}^K {{P_i}{{\left| {{\bf{f}}_{u,k}^{\rm{H}}{\bf{H}}_u^{\rm{H}}{\bf{\Theta \Phi }}{{\bf{h}}_{u,i}}} \right|}^2}}  + \sum\limits_{i = 1}^K {\kappa _{u,i}^{\rm{B}}{P_i}{{\left| {{\bf{f}}_{u,k}^{\rm{H}}{\bf{H}}_u^{\rm{H}}{\bf{\Theta \Phi }}{{\bf{h}}_{u,i}}} \right|}^2}}  + {{\left\|{ {{\bf{f}}_{u,k}^{\rm{H}}} }\right\|}^2}{\delta _{u,k}^2}} .
	\end{align}
\end{figure*}

\begin{figure*}[b]
	\setcounter{equation}{17}
	\hrulefill
	\begin{align}\label{18}
		\Gamma _{d,k,\ell}^{\rm{E}}\left( {{\bf{W}},{\bf{\Theta }}} \right) =& \sum\limits_{i = 1}^K {{{\left| {{\bf{g}}_{u,\ell}^{\rm{H}}{\bf{\Theta \Phi }}{{\bf{h}}_{u,i}}\sqrt {{P_i}} } \right|}^2}}  + \sum\limits_{i = 1}^K {\kappa_{u,i}^{\rm{B}}{{\left| {{\bf{g}}_{u,\ell}^{\rm{H}}{\bf{\Theta \Phi }}{{\bf{h}}_{u,i}}\sqrt {{P_i}} } \right|}^2}}  + \sum\limits_{i = 1,i \ne k}^K {{{\left| {{\bf{g}}_{d,\ell}^{\rm{H}}{\bf{\Theta \Phi }}{{\bf{H}}_d}{{\bf{w}}_i}} \right|}^2}}  \nonumber\\ &+  \kappa _d^{\rm{S}}{\bf{g}}_{d,\ell }^{\rm{H}}{\bf{\Theta \Phi }}{{\bf{H}}_d}\widetilde {{\rm{diag}}}\left( {\sum\limits_{i = 1}^K {{{\bf{w}}_i}} {\bf{w}}_{{i}}^{\rm{H}}} \right){\bf{H}}_d^{\rm{H}}{{\bf{\Phi }}^{\rm{H}}}{{\bf{\Theta }}^{\rm{H}}}{{\bf{g}}_{d,\ell }}.
	\end{align}
\end{figure*}
\begin{figure*}[b]
	\setcounter{equation}{18}
	\hrulefill
	\begin{align}\label{19}
		\Gamma _{u,k,\ell}^{\rm{E}}\left( {{\bf{W}},{\bf{\Theta }}} \right) =& \sum\limits_{i = 1,i \ne k}^K {{{\left| {{\bf{g}}_{u,\ell}^{\rm{H}}{\bf{\Theta \Phi }}{{\bf{h}}_{u,i}}\sqrt {{P_i}} } \right|}^2}}  + \sum\limits_{i = 1}^K {\kappa_{u,i}^{\rm{B}}{{\left| {{\bf{g}}_{u,\ell}^{\rm{H}}{\bf{\Theta \Phi }}{{\bf{h}}_{u,i}}\sqrt {{P_i}} } \right|}^2}}  + \sum\limits_{i = 1}^K {{{\left| {{\bf{g}}_{d,\ell}^{\rm{H}}{\bf{\Theta \Phi }}{{\bf{H}}_d}{{\bf{w}}_i}} \right|}^2}}  \nonumber\\&+ \kappa _d^{\rm{S}}{\bf{g}}_{d,\ell }^{\rm{H}}{\bf{\Theta \Phi }}{{\bf{H}}_d}\widetilde {{\rm{diag}}}\left( {\sum\limits_{i = 1}^K {{{\bf{w}}_i}} {\bf{w}}_{{i}}^{\rm{H}}} \right){\bf{H}}_d^{\rm{H}}{{\bf{\Phi }}^{\rm{H}}}{{\bf{\Theta }}^{\rm{H}}}{{\bf{g}}_{d,\ell }} .
	\end{align}
\end{figure*}

Then, we consider the uplink, the received signal at the BS can be written as
\begin{align}\label{YS}
{\bf{y}}_k^{\rm{S}} =& \underbrace {{\bf{H}}_u^{\rm{H}}{\bf{\Theta \Phi }}{{\bf{h}}_{u,k}}\sqrt {{P_k}}  {{{\widehat {{s}}}_{u,k}} } }_{{\rm{Desired\ signal}}}
+\underbrace {{\bf{H}}_u^{\rm{H}}{\bf{\Theta \Phi }}{{\bf{h}}_{u,k}}\sqrt {{P_k}}  { \eta _{u,k}^{\rm{B}}} }_{{\rm{Transmitter\ HWI}}}
\nonumber\\&+ \underbrace {\sum\limits_{i = 1,i \ne k}^K {{\bf{H}}_u^{\rm{H}}{\bf{\Theta \Phi }}{{\bf{h}}_{u,i}}\sqrt {{P_i}} \left( {{{\widehat {{s}}}_{u,i}} + \eta _{u,i}^{\rm{B}}} \right)} }_{{\rm{Multiuser\ interference}}} \nonumber\\&+ \underbrace {{{\bf{H}}_S}\sum\limits_{i = 1}^K {\left( {{{\bf{w}}_i}{{{s}}_{d,i}} + {\bm{\eta }}_d^{\rm{S}}} \right)} }_{{\rm{Loop-interference}}} ~+\!\!\!\!\!\! \underbrace {{\bm{\eta }}_u^{\rm{S}}}_{{\rm{Receiver\ HWI}}}   \nonumber \\
&+ \underbrace {{\bf{H}}_u^{\rm{H}}{\bf{\Theta \Phi }}{{\bf{H}}_d}\sum\limits_{i = 1}^K {\left( {{{\bf{w}}_i}{{{s}}_{d,i}} + {\bm{\eta }}_d^{\rm{S}}} \right)} }_{{\rm{Self- interfenrence}}}+ \underbrace {{{\bf{n}}^{\rm{S}}}}_{{\rm{Noise}}},\tag{12}
\end{align}
where ${\bf{H}}_{S}$ is the loop channel between the transmit antennas and receive antennas of the BS, ${\bf{n}}^{\rm{S}}$ is AWGN vector, whose elements are independent and follow the same distribution of $\mathcal{CN}(0,{\bf{\delta}} _k^2$). ${{\bm{\eta}} _u^{\rm{S}}}$ is an independent additional distortion noise term at the BS's receiver. Similar to the downlink, we express the received signal as ${\bf{ y}}_k^{\rm{S}} = \widetilde {\bf{ y}}_k^{\rm{S}} + {\bm{\eta}} _u^{\rm{S}}$, then ${{\bm{\eta}} _u^{\rm{S}}}$ follows the distribution of $\mathcal{CN}(0,\kappa _{u}^{\rm{S}}\mathbb{E}\left\{ { {\widetilde {\rm{diag}} \left\{\widetilde {\bf{ y}}_k^{\rm{S}}{{\mathop {\widetilde {\bf{y}}_k^{\rm{S}}}\nolimits^{\rm{H}} }}\right\}} } \right\})$, where $\kappa_{u}^{\rm S}$ is a proportionality coefficient that describes the severity of HWI at the BS's receiver. According to \cite{5985554} and \cite{8777303}, we assume that the BS is equipped with the SI and LI cancellation capabilities, thus LI and SI can be effectively cancelled. 
The residual noise generated in the interference cancellation process can be modeled as i.i.d. AWGN. Therefore, the total noise can be denoted as ${\bf{m}}_{\rm{S}} \buildrel \Delta \over = \left[ {{m_1},{m_2}, \ldots, {m_{{N_r}}}} \right]^T$, where ${m_i} \sim \mathcal{CN}(0,{\bf{\delta}} _{u,k}^2), i=1,2, \ldots, N_r$.

Note that the decoding matrix at the BS, which is denoted by ${{\bf{F}}}{\buildrel \Delta \over = }[{\bf{f}}_{u,1}, \ldots, {\bf{f}}_{u,k}, \ldots,{\bf{f}}_{u,K}] \in {{\mathbb C}^{N_r \times K}}$, where ${\bf{f}}_{u,k}\in {{\mathbb C}^{N_r \times 1}}$ denotes the combining vector at the BS for $\text B_{k}$. 
Therefore, the recovered signal at the BS is obtained by \cite{9318531}
\begin{align}\label{13}
{{y}}_{u,k}^{\bf{f}} 
=\ & {\bf{f}}_{u,k}^{\rm{H}}{\bf{H}}_u^{\rm{H}}{\bf{\Theta \Phi }}{{\bf{h}}_{u,k}}\sqrt {{P_k}} {\widehat {{s}}_{u,k}}+ {\bf{f}}_{u,k}^{\rm{H}}{\bm{\eta }}_u^{\rm{S}} + {\bf{f}}_{u,k}^{\rm{H}}{{\bf{m}}_{\rm{S}}} \nonumber\\&+{\bf{f}}_{u,k}^{\rm{H}}\sum\limits_{i = 1,i \ne k}^K {{\bf{H}}_u^{\rm{H}}{\bf{\Theta \Phi }}{{\bf{h}}_{u,i}}\sqrt {{P_i}} \left( {{{\widehat {{s}}}_{u,i}} + \eta _{u,i}^{\rm{B}}} \right)}\nonumber\\&+ {\bf{f}}_{u,k}^{\rm{H}}{\bf{H}}_u^{\rm{H}}{\bf{\Theta \Phi }}{{\bf{h}}_{u,k}}\sqrt {{P_k}} \eta _{u,k}^{\rm{B}} .\tag{13}
\end{align}

Accordingly, the SINR of the recovered signal at the BS is obtained by
\begin{align}\label{SNIRS}
{\rm{SINR}}_k^{\rm{S}} = \frac{{{P_k}{{\left| {{\bf{f}}_{u,k}^{\rm{H}}{\bf{H}}_u^{\rm{H}}{\bf{\Theta \Phi }}{{\bf{h}}_{u,k}}} \right|}^2}}}{{{\Gamma _k^{\rm{S}}}\left( {{\bf{W}},{\bf{\Theta }}} \right) + { {{\bf{f}}_{u,k}^{\rm{H}}{{\bm{\eta}} _u^{\rm{S}}{\mathop {\bm{\eta} _u^{\rm{S}}}\nolimits^{\rm{H}} }}{\bf{f}}_{u,k}} }}},\tag{14}
\end{align}
where ${\Gamma _{k}^{\rm S}}\left( {{\bf{W}},{\bf{\Theta }}} \right)$ is given by \eqref{15} at the bottom of the previous page.

Since the full knowledge about the eavesdroppers cannot be well acquired by the BS, we make the worst-case assumptions that the signal processing capabilities of the eavesdroppers are significantly strong and the hardware of the eavesdroppers are perfect. Thus there is no residual HWIs on the eavesdroppers\cite{7835110}.  Correspondingly, the signal eavesdropped by the $\ell$-th eavesdropper $E_\ell$ can be written as
\begin{align}\label{YE}
y_{k,\ell}^{\rm{E}} \!=& \underbrace {{\bf{g}}_{u,\ell}^{\rm{H}}{\bf{\Theta \Phi}} {{\bf{h}}_{u,k}}\sqrt {{P_k}} {{\widehat{s}}_{u,k}}}_{{\rm{Uplink\ desired\ signal}}} +\!\! \underbrace {\sum\limits_{i = 1,i \ne k}^K {{\bf{g}}_{u,\ell}^{\rm{H}}{\bf{\Theta \Phi}} {{\bf{h}}_{u,i}}\sqrt {{P_i}} {{\widehat{s}}_{u,i}}} }_{{\rm{Uplink\ multiuser\ interference}}} \nonumber\\
&+ \underbrace {\sum\limits_{i = 1}^K {{\bf{g}}_{u,\ell}^{\rm{H}}{\bf{\Theta \Phi}} {{\bf{h}}_{u,i}}\sqrt {{P_i}} \eta _{u,i}^{\rm{B}}} }_{{\rm{Uplink\ HWI}}} + \underbrace {{\bf{g}}_{d,\ell}^{\rm{H}}{\bf{\Theta \Phi}} {{\bf{H}}_d}{{\bf{w}}_k}{{{s}}_{d,k}}}_{{\rm{Downlink\ desired\ signal}}} \nonumber\\
&+ \underbrace {\sum\limits_{i = 1,i \ne k}^K {{\bf{g}}_{d,\ell}^{\rm{H}}{\bf{\Theta \Phi}} {{\bf{H}}_d}{{\bf{w}}_i}{{{s}}_{d,i}}} }_{{\rm{Downlink\ multiuser\ interference }}} +\!\! \underbrace {{{n}}_{u,\ell}^{\rm{E}}}_{{\rm{Uplink\ noise}}}\nonumber \\
&+ \underbrace { {{\bf{g}}_{d,\ell}^{\rm{H}}{\bf{\Theta \Phi}} {{\bf{H}}_d} \bm{\eta} _d^{\rm{S}}} }_{{\rm{Downlink\ HWI}}} ~+\!\!\!\! \underbrace {{{n}}_{d,\ell}^{\rm{E}}}_{{\rm{Downlink\ noise}}},\tag{16}
\end{align}
where 
${n}_{u,\ell}^{\rm{E}}\sim\mathcal{CN}(0,\mu _{{u,\ell}}^2)$ and ${n}_{d,\ell}^{\rm{E}}\sim\mathcal{CN}(0,\mu _{{d,\ell}}^2)$ denote the AWGN. The SINR at $E_\ell$ 
are respectively written as \cite{8333690}
\begin{align}\label{SINRED}
{\rm{SINR}}_{d,k,\ell}^{\rm{E}} = \frac{{{{\left| {{\bf{g}}_{d,\ell}^{\rm{H}}{\bf{\Theta \Phi}} {{\bf{H}}_d}{{\bf{w}}_k}} \right|}^2}}}{{\Gamma _{d,k,\ell}^{\rm{E}} \left( {{\bf{W}},{\bf{\Theta }}} \right) + \mu _{{\rm{d,\ell}}}^2}},\tag{17a}
\end{align}
\begin{align}\label{SINRED2}
{\rm{SINR}}_{u,k,\ell}^{\rm{E}} = \frac{{{{\left| {{\bf{g}}_{u,\ell}^{\rm{H}} {\bf{\Theta \Phi}} {{\bf{h}}_{u,k}}\sqrt {{P_k}} } \right|}^2}}}{{\Gamma _{u,k,\ell}^{\rm{E}} \left( {{\bf{W}},{\bf{\Theta }}} \right) + \mu _{{\rm{u,\ell}}}^2}},\tag{17b}
\end{align}
where $\Gamma_{d,k,\ell}^{\rm E} \left( {{\bf{W}},{\bf{\Theta }}} \right)$ and $\Gamma_{u,k,\ell}^{\rm E} \left( {{\bf{W}},{\bf{\Theta }}} \right)$ are respectively given by \eqref{18} and \eqref{19} at the bottom of this page.

Furthermore, the information rate for the $k$-th legitimate user $B_k$, the BS and the $\ell$-th eavesdropper $E_\ell$ can be written as
\begin{align}\label{RB}
{R}_k^{\rm{B}} = {\log _2}\left( {1 + {\rm{SINR}}_k^{\rm{B}}} \right),
\end{align}
\begin{align}
{R}_k^{\rm{S}} = {\log _2}\left( {1 + {\rm{SINR}}_k^{\rm{S}}} \right),
\end{align}
\begin{align}
{R}_{k,\ell}^{\rm{E}}= {\log _2}\left( {1 + {\rm{SINR}}_{d,k,\ell}^{\rm{E}}} \right)\!+\!{\log _2}\left( {1 + {\rm{SINR}}_{u,k,\ell}^{\rm{E}}} \right).
\end{align}

Since every eavesdropper has the ability to wiretap any signal of $K$ legitimate users and the BS, according to \cite{9133130}\cite{9206080}, the sum of the secrecy rate BS $\to$ $B_k$ and the secrecy rate $B_k$ $\to$ BS can be represented as
\begin{align}\label{Rsecery}
R_k^{\sec } = {\left[ {R_k^{\rm{B}}+ {R}_k^{\rm{S}}- \mathop {\max }\limits_{\forall \ell} R_{k,\ell}^{\rm{E}}} \right]^ + },\tag{23}
\end{align}
where ${\left[ x \right]^ + } = \max (0,x)$, and the SSR can be written as
\begin{align}\label{SumR}
		\setcounter{equation}{23}
C = \sum\limits_{k = 1}^K {R_k^{\sec }}.
\end{align}

\subsection{Problem Formulation}
In this work, our objective is to jointly optimize the transmit beamforming matrix $\bf{W}$ and phase shifts matrix $\bf{\Theta}$ to maximize the SSR under the transmit power constraint of the BS and the unit modulus constraint of the phase shifters. The optimization problem is formulated as
\begin{align}\label{Problem}
\mathop {\max }\limits_{{\bf{W,\Theta }}} \ &\ C\left( {{\bf{W,\Theta,\Phi }},{\bf{\mathcal {H}}}} \right)\nonumber\\
\mbox{s.t.}\quad
&\ {\rm{Tr}}\left( {{\bf{W}}{{\bf{W}}^{\rm H}}} \right) \le {P_{max }} , \\
&\left| {{\chi _m}{e^{j{\theta _m}}}} \right| = 1,0 \le {\theta _m} \le 2\pi ,1 \le m \le M \nonumber.
\end{align}

It can be seen that Problem \eqref{Problem} is a non-convex problem because of the non-convexity of the objective function and the unit modulus constraint. 
In the traditional mathematical optimization algorithm, the exhaustive algorithm can be used to obtain the optimal solution, but it is difficult to be practically implemented due to the large amount of optimization variables in our problem. We may also explore an AO algorithm to develop the suboptimal solution of the optimization problem, like
BCD algorithm in \cite{9318531}, but this also requires complex mathematical derivations, and the optimal solution is not guaranteed. Moreover, the AO algorithm cannot be directly applied to solve the considered problem due to its complicated SINR expression. Furthermore, the artificially labeled training dataset is unable
to be acquired for DL technology in dealing with this problem. The aforementioned facts motivate us to adopt the DRL-based method, which is different from other traditional mathematical methods as well as DL-based algorithms. Therefore, we utilize the advanced DRL-based algorithm to solve the joint optimization problem  without complex mathematical derivations to achieve a satisfactory transmit beamforming matrix $\bf{W}$ and phase shifts matrix $\bf{\Theta}$.

\section{DRL-Based Joint Optimization of Transmit Beamforming and Phase Shifts}
\label{section 4}
The optimization problem formulated in \eqref{Problem} is mathematically intractable as it is a non-linear and non-convex problem. Furthermore, in a practical RIS-assisted FD secure communication system with both T-HWI and RIS-HWI, the hardware quality is unknown, the capabilities of the legitimate users and  the channel quality are time-varying. The traditional optimization algorithms (such as AO and BCD) can address single time slot optimization problem, while they ignore the historical information and long-term benefits of the system and may achieve sub-optimal solutions or performances similar to the greedy-search. Therefore, it is usually not feasible to apply the traditional optimization techniques to obtain the satisfactory secure beamforming in an uncertain dynamic environment \cite{9206080}.

Model-free RL is a dynamic programming technique which learns the knowledge of the uncertain dynamic environment to deal with decision-making problems. Therefore, the Problem \eqref{Problem} is modeled as a RL problem in this paper, and a DRL-based secure beamforming algorithm is developed to jointly design the transmit and reflecting beamforming to maximize the SSR.
\subsection{Optimization Problem Transformation Based on RL}
In DRL, the RIS-aided multiuser FD secure communication system is regarded as environment, the smart controller coordinating with the BS is assumed to be an agent, which can collect CSI, such as ${\bf{H}}_u$ and all other channels. The agent interacts with the environment in discrete time slots. In time slot $t$, the agent gets the current state $s_t$ of the environment, and chooses an action $a_t$ based on the policy $\pi(s_{t},a_{t})$. After receiving the action $a_t$, the environment will update the state  to $s_{t+1}$ and feed back a reward $r_t$, which represents the performance of the action $a_t$ under current state $s_t$, and then the agent chooses action $a_{t+1}$ for the new state $s_{t+1}$. The agent exploits feedback to learn the policy to maximize cumulative rewards. The interaction process between the agent and the environment can be modeled as an MDP, which consists of a tuple denoted by $({\mathcal S}, {\mathcal A}, {\mathcal P}, r, \gamma)$. The key elements of the proposed DRL-based algorithm are defined in the following.

\textbf{1) State space}: ${s_t \in \mathcal S}$ represents the state observed from the environment at time slot \textit{t}. The state $s_t$ at time slot $t$ is composed of the rate part at the $(t-1)$-th time slot, the cascaded channel part at the $t$-th time slot, the phase noise part at the $t$-th time slot, the action at the $(t-1)$-th time slot, the transmit power of the BS and the received power of the legitimate users at the $(t-1)$-th time slot \cite{9110869}. The rate part consists of the sum rate at the legitimate users, the sum rate at the BS, the sum rate at the eavesdroppers and the SSR. The cascaded channel part contains the BS-RIS-legitimate users channel ${\bf{G}_{1d}}={\bf{h}}_{d}^{\rm{H}}{\bf{\Theta \Phi }}{{\bf{H}}_d} {{{\bf{W}}}}\in\mathbb{C}^{K \times K}$, the BS-RIS-eavesdroppers channel ${\bf{G}_{2d}}={\bf{g}}_{d}^{\rm{H}}{\bf{\Theta \Phi }}{{\bf{H}}_d} {{{\bf{W}}}}\in\mathbb{C}^{L \times K}$, the legitimate users-RIS-BS channel ${\bf{G}_{1u}}={{\bf{H}}_u^{\rm{H}}{\bf{\Theta \Phi }}{{\bf{h}}_{u}}}\in\mathbb{C}^{N_{r} \times K}$ and the legitimate users-RIS-eavesdroppers channel ${\bf{G}_{2u}}={{\bf{g}}_u^{\rm{H}}{\bf{\Theta \Phi }}{{\bf{h}}_{u}}}\in\mathbb{C}^{L \times K}$. The phase noise part consists of the random phase noise $\Delta {\theta _m}$. In order to reduce the dimension of the state space and the computational complexity, we only take the cascaded channels as the inputs instead of all the channels. Note that neural network is only capable of taking real number as input instead of complex number, so if there is complex number in state $s$, which needs to be divided into real part and imaginary part as independent input items. In total, there are $2K^2+4L{K}+2{N_r}K$ entries of state $s$ made up of the cascaded channel part. The transmit power for $B_k$ is given by ${\left\| {{{\bf{w}}_k}} \right\|^2} = {\left| {{\mathop{\rm \Re}\nolimits} \left\{ {{\bf{w}}_k^{\rm{H}}{{\bf{w}}_k}} \right\}} \right|^2} + {\left| {{\mathop{\rm \Im}\nolimits} \left\{ {{\bf{w}}_k^{\rm{H}}{{\bf{w}}_k}} \right\}} \right|^2}$. The received power for $B_k$ is given by $\left|\bf{G}_{1d,k}\right|^2=\left|\Re \left\{ \bf{G}_{1d,k}\right\}\right|^2+\left|\Im \left\{ \bf{G}_{1d,k}\right\}\right|^2$. The number of entries of the action at the $(t-1)$-th time slot is $2M+2N_tK$. To sum up, the dimension of the state space is $D_s=4+2K^2+4L{K}+2{N_r}K+4K+3M+2N_tK$.

\textbf{2) Action space}: Action space includes the transmit beamforming matrix $\bf{W}$ and the phase shifts matrix $\bm{\Theta}$. 
The diagonal elements of phase shifts matrix is regarded as entries of the action. Specifically, the real part and imaginary part, $\bm{\Theta}=\Re \left\{ \bm{\Theta}\right\}+\Im \left\{ \bm{\Theta}\right\}$ and $\bf{W}=\Re \left\{ \bf{W}\right\}+\Im \left\{ \bf{W}\right\}$ make up the action. Then, the dimension of the action space is $D_a=2M+2N_tK$. Therefore, the action vector $\bm{a}_t$ is defined as
\begin{align}
\bm{a}_t = \left[\{{\bf{w}}^{(t)}_{k}\}_{k \in K}, \{{\bm{\Theta}}^{(t)}_{m,m}\}_{m \in M} \right].
\end{align}
The action $\bm{a}_t$ will be reformed into a transmit beamforming matrix ${\bf{W}}^{(t)}$ and an RIS phase shifts matrix ${\bm{\Theta}}^{(t)}$.

\textbf{3) State transition function:} ${\mathcal P}$ is the transition probability matrix, where ${\mathcal P}(s_t,a_t,s_{t+1})\in[0,1]$ represents the probability that state $s_t$ changes to the new state $s_{t+1}$ when the agent chooses an action $a_t$.

\textbf{4) Reward function}: The objective of the optimization problem is to maximize the SSR. Thus, the reward function is defined as
\begin{align}\label{Reward}
r_t =C^{(t)}.
\end{align}

\textbf{5) Discount factor:} $\gamma\in[0,1]$ is applied to discount future rewards and represents the uncertainty of future rewards.

\textbf{6) Policy:} the policy $\pi(s_t,a_t)$ denotes probability distribution of choosing an action $a_t$ over the state $s_t$, which satisfies $\sum\nolimits_{{a_t} \in \mathcal{A}} {\pi \left( {{s_t},{a_t}} \right)}  = 1$.

\subsection{DDPG-Based Joint Transmit and Reflecting Beamforming Optimization}
\label{Algorithm Description}

In DRL, whether transition probability is known or not is a big difference between model-based and model-free algorithms. The proposed multiuser FD PLS problem is built as an MDP model, once the transition probability set is given, the problem can be effectively solved by dynamic programming techniques \cite{8496766}. However, this is not the case, as the transition probability set is difficult to obtain in most cases. Therefore, some standard RL algorithms, such as Q-learning, are considered to be instrumental in solving the model-free problems, in which the transition probability is not required.
However, Q-learning is unable to deal with the high-dimensional input problem and the continuous state-action space. To put it in practical terms, the channel matrix and beamforming matrix in the considered RIS-aided FD secure communication system, which are difficult to enumerate or even to discretize.

In order to overcome the shortcomings of tabular RL algorithms as well as to tackle the continuous action and state space,
we proposed DDPG-based algorithm, which is based on actor-critic framework and shown in Fig. \ref{DDPGFramework}, to jointly design the BS transmit beamforming and the phase shifts for the RIS-aided FD secure communication system. There are four neural networks in the proposed DDPG-based secure beamforming algorithm, named actor network,  target actor network, critic network and target critic network, the function of each network is listed as below \cite{9322175}:

\begin{figure}[t]
	\hspace{-0.25cm}
	\includegraphics[scale=0.96]{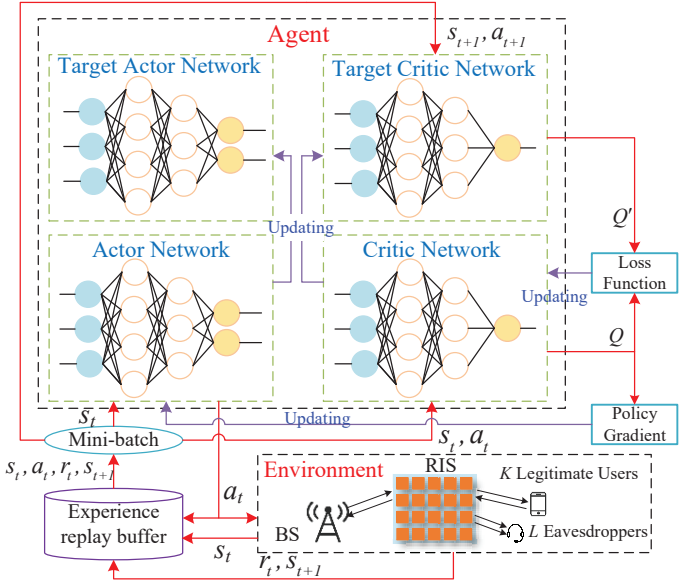}
	\caption{The framework of the proposed DDPG algorithm for RIS-aided multiuser FD two-way communication system with multi-eavesdropper.}
	\label{DDPGFramework}
\end{figure}

\textbf{1) Actor network}: It is also known as policy network, takes state $s$ as input and outputs an continuous action $a$, update network parameter $\theta_\mu$. The actor network is trained by the equation \eqref{Optimal action}, which aims to maximize the state-value function. The update on the actor network is expressed as
	\begin{align}\label{Actor update}
		{\theta _\mu } = {\theta _\mu } - {\alpha _\mu }\nabla Q\left( {{\theta _{ct}}|s,a} \right)\nabla \pi \left( {{\theta _\mu }|s} \right),
	\end{align}
	where $\alpha_\mu$ is the learning rate for the actor network, $\nabla\left(\cdot \right)$ denotes the gradient and $\theta_{ct}$ is the parameter of the target critic network.
	
\textbf{2) Target actor network}: It is parameterized by $\theta_{\mu_t}$, and $\theta_{\mu_t}$ is periodically soft updated by
	\begin{align}\label{update target actor}
		{\theta _\mu }_t \leftarrow {\beta _\mu }{\theta _\mu } + \left( {1 - {\beta _\mu }} \right){\theta _\mu }_t,
	\end{align}
where ${\beta _\mu }$ is the learning rate for the target actor network.

\textbf{3) Critic network}: It is also called Q-network, parameterized by $\theta_c$, takes state $s$ and action $a$ as input and outputs the Q-function defined in \eqref{Q funtion}. The loss function $L(\theta)$ is a difference between the output Q-value of the target neural network and the predict neural network, therefore, the loss function is defined by
	\begin{align}\label{Loss function}
		L\left( {{\theta _{c}}} \right) = {\left( {{y} - Q\left( {{\theta _{c}|s_t,a_t}} \right)} \right)^2},
	\end{align}
	where ${y}$ is the target value, which is estimated by
	\begin{align}\label{Yi}
		{y} = {r_{t + 1}} + \gamma \mathop {\max }\limits_{{a_{t + 1}}} Q\left( {{\theta _{c}}|{s_{t + 1}},{a_{t + 1}}} \right).
	\end{align}
	The gradient update of the loss function \eqref{Loss function} with respect to $\theta_c$ is updated by
	\begin{align}\label{Loss function update}
		{\theta _c} = {\theta _c} - {\alpha _c}\nabla L\left( {{\theta _c}} \right),
	\end{align}
	where $\alpha_c$ is the learning rate for the critic network.
	
\textbf{4) Target critic network}: It is parameterized by $\theta_{ct}$, and outputs ${Q_t}\left( {{\theta _{ct}}|{s_{t + 1}},{a_{t + 1}}} \right)$, $\theta_{ct}$ is soft updated by
	\begin{align}\label{update target critic}
		{\theta _c }_t \leftarrow {\beta _c }{\theta _c } + \left( {1 - {\beta _c }} \right){\theta _c }_t,
	\end{align}
where ${\beta _c }$ is the learning rate for the target critic network.

In the process of interaction between agent and environment, the goal of the agent is to search the optimal policy $\pi^*$ to maximize the long-term expected discounted reward, and the cumulative discounted reward function is defined as
\begin{align}\label{reward function}
	{R_t} = \sum\limits_{\tau  = 0}^\infty  {{\gamma _\tau }{r_{t + \tau  + 1}}} ,
\end{align}
where $\tau$ is the number of time slots. When given a certain policy $\pi_t$ and a state-action pair $(s_t,a_t)$, the cumulative discounted reward can be approximately calculated by
\begin{align}\label{Q funtion}
	{Q_\pi }\left( {{s_t},{a_t}} \right) = \mathbb{E}_\pi \left[ {{R_t}|{s_t} = s,{a_t} = a} \right],
\end{align}
which is called the action-value function and is also known as the Q-function. It can be updated by Bellman expectation equation\cite{9206080}.
The optimal Q-function in \eqref{Q funtion} can be solved by Bellman optimality equation as follows:
\begin{align}\label{Optimal Bellman}
	{Q^ * }\left( {{s_t},{a_t}} \right) = {r_t} + \gamma \mathop {\max }\limits_{{a_{t + 1}}\in \mathcal{A}} {Q^ * }\left( {{s_{t + 1}},{a_{t + 1}}} \right),
\end{align}
the corresponding optimal action $a^*$ 
can be obtained by
\begin{align}\label{Optimal action}
	{a^*} = \arg \mathop {\max }\limits_{a \in \mathcal{A}} Q^*\left( {s,a} \right).
\end{align}

The details of the proposed DRL-based method for joint optimization of the transmit beamforming and the phase shifts are shown in {\bf{Algorithm \ref{DDPG Algorithm}}}. The algorithm runs over $N^{epi}$ episodes and each episode iterates $T$ steps. At the beginning of each episode, the parameters of the four networks are initialized, including $\theta_\mu$ in the actor network, $\theta_{\mu t}$ in the target actor network, $\theta_c$ in the critic network and $\theta_{ct}$ in the target critic network, all of which are uniformly distributed. In addition, the experience replay memory $\mathcal{M}$ with size $\mathcal{D}$ will be emptied. For the sake of encouraging the agent to fully explore the environment, exploration noise $\mathcal{N}$ is added to the output of the actor network. It implies that the action to be chosen for state $s$ is given by
$\widetilde{a_t}=a_t+\mathcal{N}$,
where $a_t$ is the output of the actor network at time slot $t$, and the exploration noise $\mathcal{N}$ can be chosen appropriately to suit the environment and will decrease with the increase of the number of iterations. We adopt the identity matrix to initialize the action $a_0$, including the transmit beamforming matrix $\bf{W}_0$ and the phase shifts matrix $\bm{\Theta}_0$. In addition, all the channels and the exploration noise $\mathcal{N}$ will be initialized at the beginning of each episode.

\begin{algorithm}[t] 
\caption{DDPG-based joint optimization of transmit beamforming and phase shifts} 
\label{DDPG Algorithm} 
\textbf{Input}: All the channels, ${{\bf{H}}_{d}}$, ${{\bf{H}}_{u}}$, ${{\bf{h}}_{d,k}}$, ${{\bf{h}}_{u,k}}$, ${{\bf{g}}_{d,\ell}} $, ${{\bf{g}}_{u,\ell}}$; the coefficients of HWI, $\kappa _{u, k}^{\rm{B}}$, $\kappa _{u}^{\rm{S}}$, $\kappa_{d, k}^{\rm{B}}$, $\kappa _{d}^{\rm{S}}$.

\textbf{Output}: The optimal phase shifts matrix $\bm{\Theta}^*$, the optimal transmit beamforming matrix $\bf{W}^*$, the maximized SSR $C^*$ under current channel state.

\textbf{Initialize}: actor network parameter $\theta_\mu$, target actor network parameter $\theta_{\mu t}$, critic network parameter $\theta_c$, target critic network parameter $\theta_{ct}$, learning rate $\alpha_\mu$, $\alpha_c$, $\beta_\mu$ and $\beta_c$, mini-batch size $H$, experience replay memory $\mathcal{M}$ with size $D$.
\begin{algorithmic}[1]
\For{episode = $1, 2 , \ldots, N^{epi}$}
      \State Initialize the exploration noise $\mathcal{N}$;
      \State Initialize the phase shifts matrix $\bm{\Theta}_0$ and the beam-
      \Statex ~~~~~forming matrix $\bf{W}_0$;
      \State Obtain the initial state $s_0$;
      \For {time slot $t = 1, 2, \ldots, T$}
            \State Obtain action $a_t$ from the actor network;
            \State Add exploration noise to $a_t$ as $\widetilde{a_t}=a_t+\mathcal{N}$;
            \State Calculate the instant reward $r_t$ using \eqref{Reward};
            \State Observe the new state $s_{t+1}$;
            \State Store transition $\{ s_t, a_t, r_t, s_{t+1}\}$ into $\mathcal{M}$;
            \State Sample a $H$ mini-batch transitions from $\mathcal{M}$;
            \State Set target function $y$ according to \eqref{Yi};
            \State Minimize the loss function $L\left( {{\theta _{c}}} \right)$ by equation \eqref{Loss function};
            \State Update the actor network ${\theta _\mu }$ by \eqref{Actor update} and the critic 
            \Statex~~~~~~~~~network ${\theta _c }$ by \eqref{Loss function update};
            \State Update the target actor network ${\theta _{\mu t} }$ by \eqref{update target actor} and 
            \Statex ~~~~~~~~~the target critic network ${\theta _{ct} }$ by \eqref{update target critic};
            \State Update the sate $s_t=s_{t+1}$;
            \State Reduce the exploration noise $\mathcal{N}$.
      \EndFor
\EndFor

\end{algorithmic}
\end{algorithm}

All the proposed neural network structures are based on fully connected layer. The actor network structure is composed of five layers, including an input layer, an output layer and three hidden layers. The critic network shares almost the same network structure as the actor network. The input and output dimensions of the actor network are equal to state and action dimensions, respectively, and the input dimension of critic network is equal to the sum of action and state dimensions.
Due to the negative inputs, the activation function $tanh$ is utilized for the network. The optimizer applied for the network is Adam, which is with adaptive learning rate. More details are shown in TABLE \refeq{tab:2}. Furthermore, the correlation between input data will adversely affect the learning of neural networks. In order to reduce the correlation among the data in the state, the state needs to go through the whitening process before being input into the actor network and critic network \cite{9110869}, which is conducive to the learning of neural network.

Note that the transmit beamforming matrix $\bf{W}$ and the phase shifts matrix $\bm{\Theta}$ need to meet the power constraint defined
in \eqref{8} and the unit modulus constraints, respectively. In order to satisfy these two constraints, we add a batch normalization layer after the output layer of actor network, where ${\rm{Tr}}\left( {{\bf{W}}{{\bf{W}}^{\rm H}}} \right) \le {P_{max }}$ and the phase shifts satisfy $\left| {{\chi _m}{e^{j{\theta _m}}}} \right| = 1$, the later constraint means the transmission signal will change the direction without power loss when going through RIS. Specifically, to satisfy the BS maximum power constraint, the following projection operator is applied.
\begin{equation}
	\prod {\left\{ {\bf{W}} \right\}}  = \left\{ {\begin{array}{*{20}{c}}
			~{{\bf{W}},~~~~~~~{\rm{Tr}}\left( {{\bf{W}}{{\bf{W}}^{\rm H}}} \right) \le {P_{\max}}},\\
			\!\!\!\!\!\!\!\!\!\!\!\!\!\!\!\!\!\!\!\!\!\!\!\!\!\!\!\!\!\!{\frac{{\sqrt {{P_{\max}}} }}{{\left\| {\bf{W}} \right\|}_F}{\bf{W}},~~{\text{otherwise}}}.
	\end{array}} \right.\label{Normalize_W}
\end{equation}
After the normalization operation, we can obtain and reformulate the normalized transmit beamforming ${{\bf{W}}_{{\rm{nor}}}} = \prod {\left\{ {\bf{W}} \right\}}$, and then ${\rm{Tr}}({{\bf{W}}_{{\rm{nor}}}}{{\bf{W}}^{\rm H}_{{\rm{nor}}}})\le {P_{\max}}$, which always satisfies the BS maximum power constraint in Problem \eqref{Problem}.

\begin{figure}
	\centering
	\includegraphics[scale=0.75]{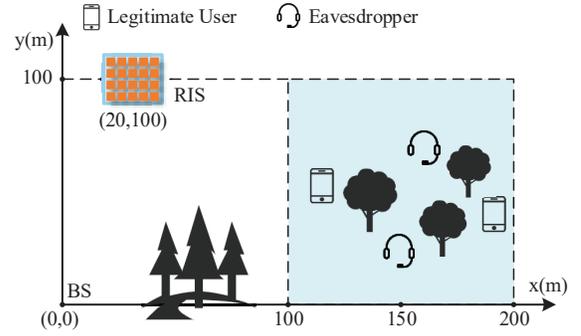}
	\caption{Simulation setup for RIS-aided multiuser FD two-way communication system with multi-eavesdropper.}
	\label{Simulation model}
\end{figure}

\renewcommand\arraystretch{1.2}  
\renewcommand\tabcolsep{3.0 pt} 
\begin{table}
		\caption{Parameters for Simulation}
		\label{tab:1}       
		\begin{center}
			\begin{tabular}{|c|c|c|}
				\hline
				{\bfseries Parameters}  & \bfseries {Description} & \bfseries settings\\
				\hline
				{\bfseries $P_{max}$}  &  transmit power at BS& $10$ dBm $\sim$ 40 dBm \\
				\hline
				{\bfseries $P_k$}  &  transmit power at users & 100 mW \\
				\hline
				{\bfseries $\alpha$}  &  path loss exponent & 2.0 \\
				\hline
				{\bfseries $\varepsilon_{\bf{h}}$}  & Rician factor & 10  \\
				\hline
				{\bfseries $BW$}  & channel bandwidth & 10 MHz \\
				\hline
				{\bfseries $\delta^2$}  & noise power density & $-174$ dBm/Hz \\
				\hline
				{\bfseries $\rho_S$}  &  SI coefficient & 1 \\
				\hline
			\end{tabular}
		\end{center}
\end{table}

Our objective is to take advantage of DRL to find satisfactory $\bf{W}$ and $\bm{\Theta}$ by maximizing the SSR under given a particular CSI, according to \cite{9110869}, rather than offline training and online deployment such as those in \cite{9206080}. 
In the presented DRL-based algorithm, each CSI is utilized to construct the state, then the DDPG algorithm is run to find the optimal solutions. In this paper, the optimal transmit beamforming matrix $\bf{W}^*$ and the phase shifts matrix $\bm{\Theta}^*$ are obtained by the action with the largest instant reward.

\section{Simulation Results}
\label{section 5}
In this section, extensive simulation results are presented to evaluate the performance of the DRL-based algorithm in the RIS-aided multiuser FD secure communication system with multiple eavesdroppers.
\subsection{Simulation Setup}

\begin{figure}[t]
	\centering
	\includegraphics[scale=0.6]{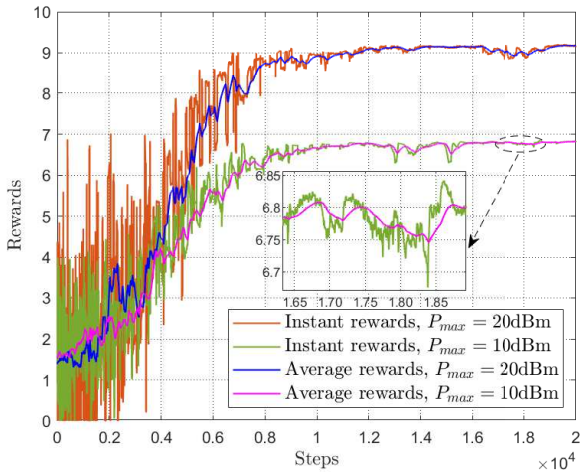}
	\caption{Average reward and instant reward performance versus time steps at $P_{max}=20$ dBm and $P_{max}=10$ dBm, respectively.}
	\label{Convergence_20dB_10dB}
\end{figure}
\begin{table}[t]
		\renewcommand\arraystretch{1.2}  
		\renewcommand\tabcolsep{6.0 pt} 
		\caption{Settings for DDPG Networks}
		\label{tab:2}       
		\begin{center}
			\begin{tabular}{|c|c|c|c|}
				\hline
				{\bfseries Network}  & \bfseries {Layer} & \bfseries Neurons &{\bfseries Activation} \\
				\hline
				{\bfseries }  &  Input& $D_s + D_a$ &/\\
				\cline {2-4}
				&  Hidden 1 & 128& ReLU \\
				\cline {2-4}
				{\bfseries critic network}  &  Hidden 2 & 128&ReLU\\
				\cline {2-4}
				{\bfseries }  & Output & $1$ &ReLU\\
				\hline
				{\bfseries }  & Input & $D_s$ &/\\
				\cline {2-4}
				{\bfseries }  &  Hidden 1 & 128 &ReLU\\
				\cline {2-4}
				{\bfseries actor network}  &  Hidden 2 & 128 &ReLU\\
				\cline {2-4}
				{\bfseries }  &  Hidden 3 & 128 &ReLU\\
				\cline {2-4}
				{\bfseries }  &  Output & $D_a$ &tanh\\
				\hline
			\end{tabular}
		\end{center}
\end{table}

Fig. \ref{Simulation model} depicts a two-dimensional plane of the proposed RIS-assisted FD secure communication system, the default unit in the figure is meter (m). As shown in the figure, $K$ legitimate users and $L$ eavesdroppers are randomly distributed in the light blue square in the right half of the figure, with the size of 100 m × 100 m. The BS and RIS are located at (0,0) and (20,100), respectively.

Unless stated otherwise, the following parameters are employed in the simulations. The numbers of antennas at the BS are set to $N_t=N_r=4$, the number of the RIS reflecting elements is set to $M=8$, the number of the legitimate users is set to $K=2$ and the number of the eavesdroppers is set to $L=2$. 
The large-scale path loss 
in equation \eqref{1} is defined by $PL=PL_0-10\alpha\log_{10}(d/d_0)$, where $PL_0=-30$ dB is the path loss at 
$d_0=1~m$, $d$ is the distance between the transmitter and the receiver.
All channels follow the Rician distribution. 
Let ${\bf{\delta}} _{u,k}^2=1.1{\bf{\delta}} _{k}^2$, $\sigma _{d,k}^2=1.1\sigma _{k}^2$, the coefficients of HWI $\kappa _{u,k}^{\rm{B}}=\kappa _{u}^{\rm{S}}=\kappa_{d,k}^{\rm{B}}=\kappa _{d}^{\rm{S}}=\kappa=0.01$, the random phase errors are uniformly distributed within $\left[ {{{ - \pi } \mathord{\left/
 {\vphantom {{ - \pi } {2,{\pi  \mathord{\left/
 {\vphantom {\pi  2}} \right.
 \kern-\nulldelimiterspace} 2}}}} \right.
 \kern-\nulldelimiterspace} {2,{\pi  \mathord{\left/
 {\vphantom {\pi  2}} \right.
 \kern-\nulldelimiterspace} 2}}}} \right]$.
The other parameters are given in TABLE \ref{tab:1}.

In the proposed DRL algorithm, the discount factor is set to $\gamma=0.9$, the learning rate is set to $\alpha_\mu=0.0005, \alpha_c=0.001$, the batch size of the mini-batch is set to $H=128$, the experience replay capacity is set to $D=100000$, the number of steps in each episode is set to $T=20000$. Other neural network parameter settings, such as activation function, are given in TABLE \ref{tab:2}.

\subsection{Convergence}

\begin{figure}[t]
	\centering
	\includegraphics[scale=0.6]{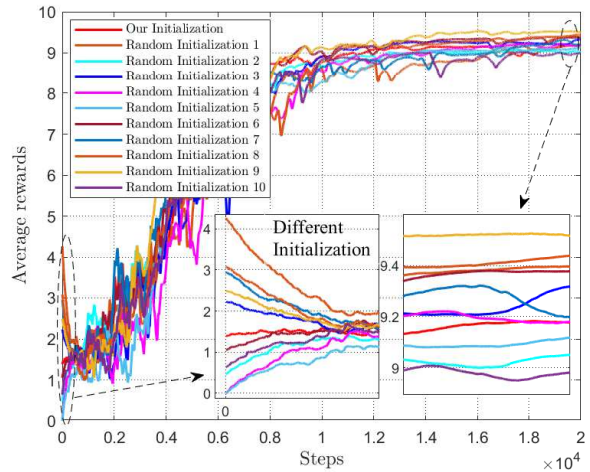}
	\caption{Average reward performance for 10 random initializations.}
	\label{Impact_of_Initialization}
\end{figure}
Before showing the simulation results, the convergence performance of the presented DDPG algorithm 
is verified. Fig. \ref{Convergence_20dB_10dB}. 
shows the convergence behavior of the proposed method.
In order to better demonstrate the convergence performance of the proposed algorithm, instant reward and average reward are considered, 
among which the average reward is defined as follows\cite{9667526}:
\begin{align}\label{Average reward defined}
	r^{{\text{Average}}}_{t + 1} = \varpi   r^{{\text{Average}}}_t + \left( {1 - \varpi} \right)r^{{\text{Instant}}}_{t + 1},
\end{align}
where $\varpi$ is the smoothing factor, $t$ is the time step. In this paper, the reward function represents the optimization
objective, and our objective is to maximize the SSR. Thus, the reward function is defined as the SSR. 
The convergence behavior of the DDPG algorithm under different BS transmit power is shown in Fig. \ref{Convergence_20dB_10dB}. It is obvious from the figure that both the instant rewards and the average rewards will converge with the increase of time step $t$. This result indicates that the DRL-based algorithm explores the environment through interactions between agent and environment, then learns from the exploration experience to converge at about $t = 8000$ steps and obtains a satisfactory solution.

\begin{figure}[t]
	\centering
	\includegraphics[scale=0.6]{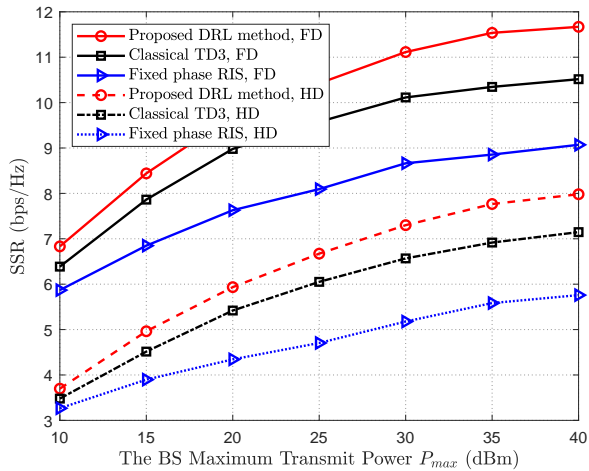}
	\caption{SSR versus $P_{max}$ with $K=2$, $L=2$, $\kappa=0.01$ under different schemes.}
	\label{benchmark}
\end{figure}

Consider the non-convexity of Problem \eqref{Problem}, different initial points may lead to different local optimal solutions of the developed algorithm. Fig. \ref{Impact_of_Initialization} investigates the impact of different initializations on the performance of proposed algorithm. The initialization strategy of {\bf{Algorithm \ref{DDPG Algorithm}}} is that the identity matrix is adopted to initialize the action $a_0$, including the transmit beamforming matrix $\bf{W}_0$ and the phase shifts matrix $\bm{\Theta}_0$. ``Random initialization" denotes that the initial point of {\bf{Algorithm \ref{DDPG Algorithm}}} is randomly selected under the constraints in Problem \eqref{Problem}. As can be seen from the figure that although the initial points of each curve are different, the convergence points are almost the same. Firstly, it shows that the initialization strategy adopted by the proposed algorithm is a good choice. In addition, it is also proved that the proposed algorithm is robust to the initial points. In other words, the proposed DRL-based algorithm can converge to a satisfactory solution that is almost the same even though the initial points are different.

\begin{figure}[t]
	\centering
	\includegraphics[scale=0.6]{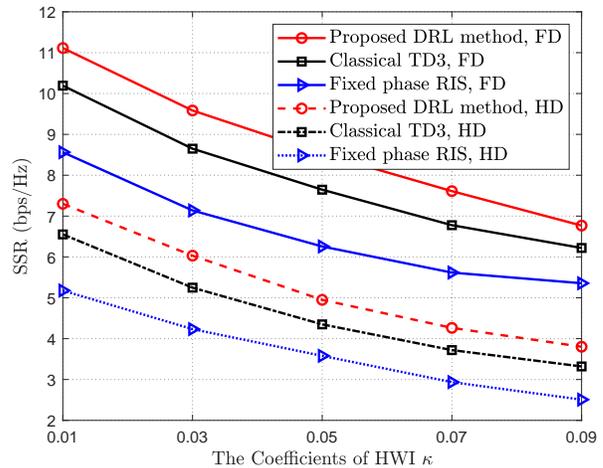}
	\caption{SSR versus $\kappa$ with $K=2$, $L=2$, $P_{max}=30$ dBm under different schemes.}
	\label{ImpactOf_kappa}
\end{figure}

\subsection{Comparisons with Benchmarks}

Fig. \ref{benchmark} illustrates the SSR versus $P_{max}$ under several different schemes. We consider six schemes, ``Proposed DRL method, FD'' and ``Proposed DRL method, HD" denote {\bf{Algorithm \ref{DDPG Algorithm}}} with FD and HD, i.e., $P_k = 0$ mW. For the sake of comparison, classical TD3 \cite{TD3} is applied to the joint design of transmit beamforming and RIS phase shifts as the performance benchmark, which are denoted by ``Classical TD3, FD'' and ``Classical TD3, HD". Moreover, ``Fixed phase RIS, FD'' and ``Fixed phase RIS, HD" denote that the phase shift for each reflecting unit is fixed while the transmit beamforming is optimized by {\bf{Algorithm \ref{DDPG Algorithm}}}, which serves as another performance benchmark. From the figure, firstly, the proposed DRL scheme achieves much higher SSR than both classical TD3 and fixed phase RIS scheme. Furthermore, with the increase of BS maximum transmit power $P_{max}$, the SSRs of all schemes increase as expected, either in FD mode or HD mode. 

Fig. \ref{ImpactOf_kappa} shows the SSR performance versus $\kappa$ under different schemes. It is obvious from the figure that the performance of various schemes deteriorates with the increase of $\kappa$, owing to the fact that adverse effect of hardware impairments is more prominent. In addition, it can be found that the SSR in FD mode decreases faster than the SSR in HD mode with the increase of $\kappa$. This is because the distortion noise power in FD mode increases much more than the distortion noise power in HD mode with the increase of $\kappa$. In other words, for a relatively small $\kappa$, a greater performance gain will be achieved by FD mode in contract with HD mode. 

\subsection{Impact of the System Settings}

\begin{figure}[t]
	\centering
	\includegraphics[scale=0.6]{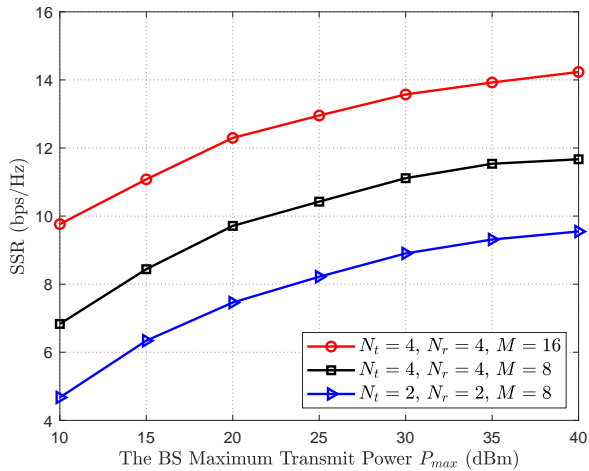}
	\caption{SSR as a function of $P_{max}$ under different system settings.}
	\label{Impact_of_System settings}
\end{figure}

In order to further verify the effectiveness of our proposed DRL-based method, we evaluate the performance under various system settings. We consider three cases, i.e., $\{N_t = 2, N_r = 2, M = 8\}$, $\{N_t = 4, N_r = 4, M = 8\}$ and $\{N_t = 4, N_r = 4, M = 16\}$, which are shown in Fig. \ref{Impact_of_System settings}. From this figure, we see that, the SSR increases
with $P_{max}$. As more transmit power is allocated to BS, higher SSR can be obtained. This observation is consistent with that of conventional secure systems. In addition, larger size of RIS and more antennas at the BS lead to higher SSR.

\begin{figure}[t]
	\centering
	\includegraphics[scale=0.6]{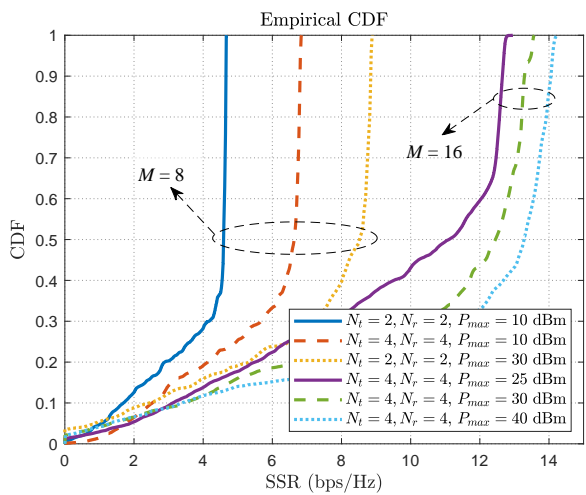}
	\caption{CDF of SSR for various system settings.}
	\label{CDF of_System settings}
\end{figure}

\begin{figure}
	\centering
	\includegraphics[scale=0.6]{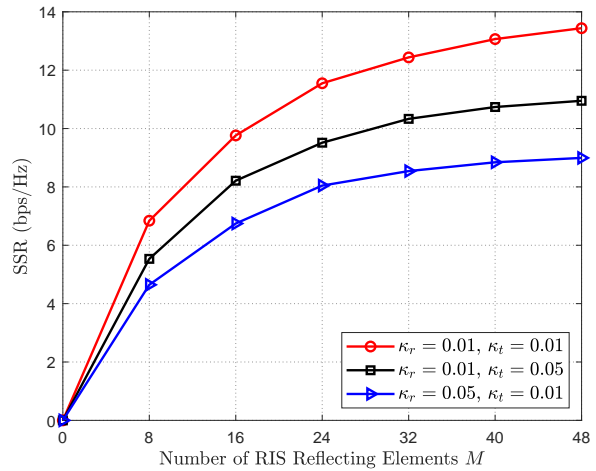}
	\caption{SSR versus $M$ with $P_{max}=10$ dBm, $K=2$, $L=2$ and different values of $\kappa$.}
	\label{Impact_of_RIS_Elements}
\end{figure}

Furthermore, we also compare the cumulative distribution function (CDF) of SSRs under different system settings, which are shown in Fig. \ref{CDF of_System settings}. It is obvious from the figure that the CDF curves demonstrate the observations from Fig. \ref{Impact_of_System settings}, where the SSR improves with the increase of the BS transmit power $P_{max}$, the number of antennas at the BS and the number of RIS reflecting elements $M$. The performance gain of the proposed DRL-based algorithms is stable in the light of the CDF curves, which means that the presented algorithm is robust to the system settings and can perform well with high probability. On the other hand, the CDF curves demonstrate the convergence of proposed algorithm shown in Fig. \ref{Convergence_20dB_10dB}. For example, focus on the curve with $\{N_t = 2, N_r = 2, P_{\max} = 10~{\rm{dBm}}, M = 8\}$, there is a $40\%$ probability that the abscissa of this curve is less than 4.5, and $60\%$ is stable at about 4.5. This proves that the agent in proposed DRL-based method explores in the environment in the first 8000 steps, then learns from the exploration experience to converge and obtains a satisfactory solution in 8000-20000 steps.

\subsection{Impact of Number of Reflecting Elements at RIS}
Fig. \ref{Impact_of_RIS_Elements} depicts the impact of the number of the reflecting elements $M$ on the SSR performance with $K=2,L=2$, $\kappa_r = \kappa^{\rm {B}}_{d,k} = \kappa^{\rm {S}}_u = \{0.01, 0.05\}$ and $\kappa_t = \kappa^{\rm {B}}_{u,k} = \kappa^{\rm {S}}_d = \{0.01, 0.05\}$ when ${P_{max} = 10}$ dBm. On the one hand, it can be seen that increasing the number of RIS's reflecting elements will increase the SSR of the system. However, the performance gain decreases while the level of the HWI increases. On the other hand, the SSR in the system with $\{\kappa_r = 0.01, \kappa_t = 0.05\}$ is much higher than the SSR in the system with $\{\kappa_r = 0.05, \kappa_t = 0.01\}$. This observation means that the HWI on receivers has a greater negative impact on the SSR than that of the HWI on the transmitters. This is because the proportion of the receive distortion noise power in the total received noise is much larger than that in the transmit distortion noise power.

\subsection{Impact of the Path Loss Exponent}

\begin{figure}
	\centering
	\includegraphics[scale=0.6]{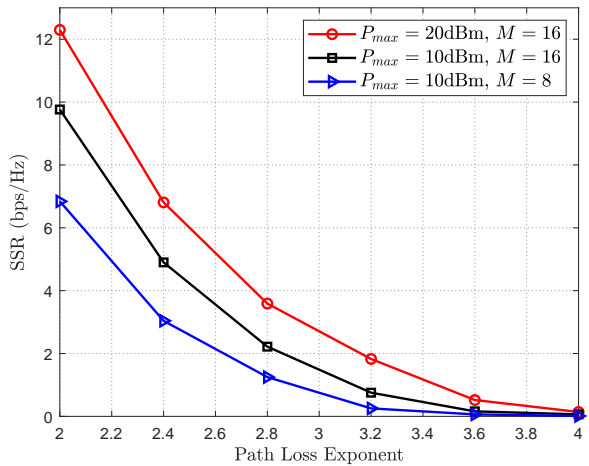}
	\caption{SSR versus pass loss exponent $\alpha$ with $P_{max}=\{10, 20\}$ dBm and $M = \{8, 16\}$.}
	\label{ImpactOf_PassLoss}
\end{figure}

In the aforementioned simulations, the path loss exponent is set as 2.0 due to the assumption that the RIS is deployed in an unobstructed location so that the link of BS-RIS-Users is established. However, in some practical communication scenarios, this ideal place may not exist. Therefore, the impact of large-scale fading channel on the performance of secure system is investigated. Fig. \ref{ImpactOf_PassLoss} shows the SSR versus the path loss exponent. The SSR obtained by the proposed DRL algorithm decreases with the increase of $\alpha$ as expected. This is because larger-scale fading will lead to a weaker signal reflected from the RIS, which weakens the benefits of RIS. Fortunately, this provides practical engineering guidance that RIS should be carefully deployed in a place with less obstacles in the legitimate link and more obstacles in the eavesdropping link to improve the system's secrecy performance.

\begin{figure}
	\centering
	\includegraphics[scale=0.6]{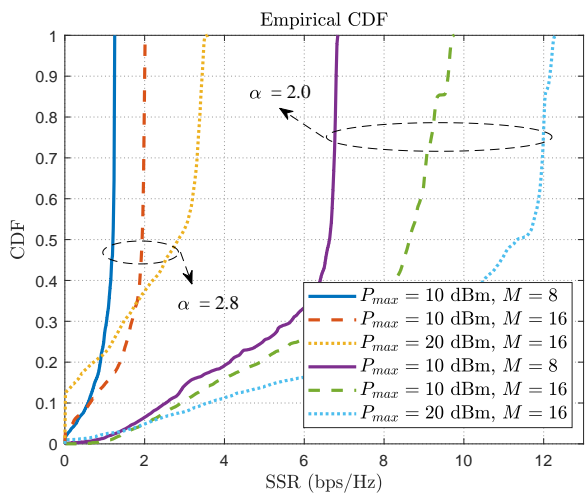}
	\caption{CDF of SSR for various $\alpha$, $P_{max}$ as well as $M$.}
	\label{CDF_of_PassLoss}
\end{figure}

Furthermore, we also compare the CDF of SSR for various $\alpha$, $P_{max}$ as well as $M$, i.e., $\alpha = \{2.0, 2.8\}$, $P_{max} = \{10, 20\}$ dBm and $M = \{8, 16\}$, which are shown in Fig. \ref{CDF_of_PassLoss}. It is obvious from the figure that the CDF curves are consistent with the results in Fig. \ref{ImpactOf_PassLoss}. In addition, the performance gain obtained by the proposed DRL method is stable according to the CDF curves.



\subsection{Impact of Hyper-Parameters of DRL}
\begin{figure}
	\centering
	\includegraphics[scale=0.6]{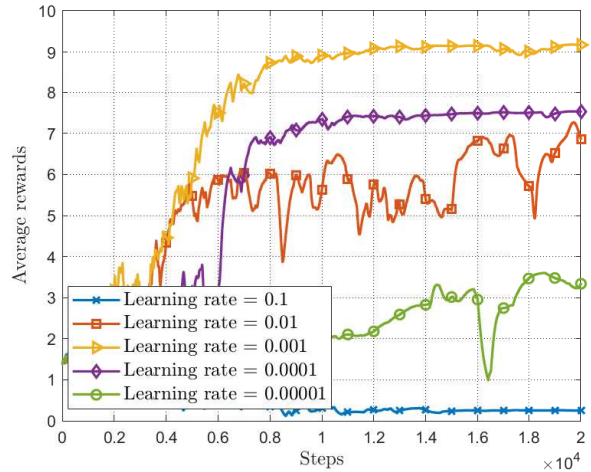}
	\caption{Average rewards performance versus time steps under different learning rates, i.e.,\{0.1, 0.01, 0.001, 0.0001, 0.00001\}.}
	\label{Impact_of_LearningRate}
\end{figure}

\begin{figure}
	\centering
	\includegraphics[scale=0.6]{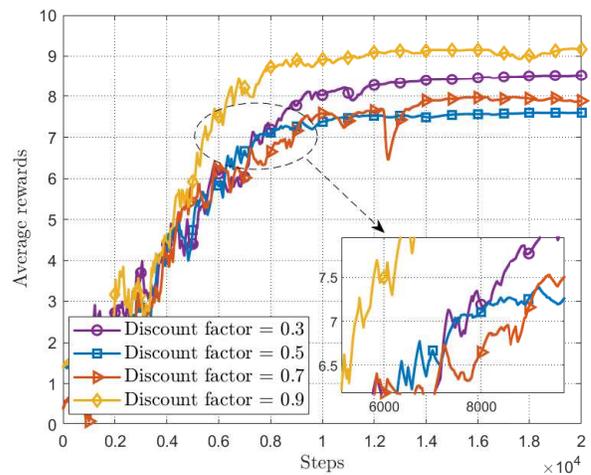}
	\caption{Average rewards performance versus time steps under different discount factors, i.e.,\{0.3, 0.5, 0.7, 0.9\}.}
	\label{Impact_of_DiscountFactor}
\end{figure}

Network parameters play an important role in DRL, which determine the convergence speed and efficiency of learning. By choosing appropriate network parameters, the DRL can achieve the desired results. Here, we take the learning rate and the discount factor as examples to illustrate that selecting network parameter appropriately is crucial to DRL. Fig. \ref{Impact_of_LearningRate} exhibits the average rewards versus time steps under different learning rates, i.e., $\{0.1, 0.01, 0.001, 0.0001, 0.00001\}$. It is seen from the figure that learning rates with different levels have a great impact on the performance of DRL. Specifically, compared with other learning rates in the figure, the DRL with learning rate $\alpha_c=0.001$ achieves the best performance. If the learning rate is too large, it will increase the oscillation, which will lead to poor performance or even decline in performance. If the learning rate is too small, it will take a long time to achieve convergence or it is difficult to converge to a good result. Therefore, the learning rate should be set appropriately, neither too large nor too small. Hence, in the proposed DRL algorithm, the learning rate is set to 0.001, which can achieve better performance. Fig. \ref{Impact_of_DiscountFactor} compares the average rewards performance versus time steps under different discount factors, i.e.,\{0.3, 0.5, 0.7, 0.9\}. It shares similar conclusions with the learning rate, but it has less effect on the performance and convergence speed of DRL. It can be seen from the figure that the discount factor with 0.9 achieves the best performance.

Finally, we can conclude that DRL is a complex dynamic learning process, and the selection of the network parameters determines its performance, convergence speed and efficiency of learning. These parameters include not only learning rate, discount factor, delaying rate but also activation function, etc. Choosing appropriate parameters can improve the performance of DRL-based algorithm and accelerate the convergence speed of neural network.
\section{Conclusion}
\label{section 6}
In this paper, we studied an RIS-assisted FD secure communication system in the presence of HWI. A SSR maximization problem was formulated by jointly optimizing both the transmit beamforming at the BS and the phase shifts at the RIS. To tackle this mathematically intractable NP-hard problem, a DRL-based algorithm was proposed to obtain the satisfactory solution. Moreover, the proposed DRL method had a standard framework and low complexity in implementation, without explicit mathematical calculation. Therefore, it was conveniently deployed to communication systems with different settings. Extensive simulation results confirmed the effectiveness of the developed DRL-based algorithm in improving the SSR. It was also found that appropriate neural network parameters can improve the performance of DRL-based algorithm and accelerate the convergence speed of the neural network.

%



\bibliographystyle{IEEEtran}
\bibliography{IEEEabrv,Refer}
\vspace{-0.8cm}
\begin{IEEEbiography}[{\includegraphics[width=1in,height=1.25in,clip,keepaspectratio]{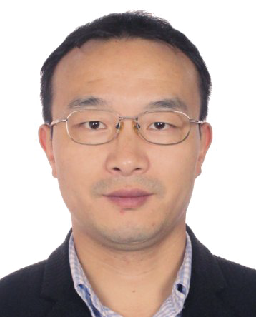}}]{Zhangjie Peng}
	received the B.S. degree from Southwest Jiaotong University, Chengdu, China, in 2004, and the M.S. and Ph.D. degrees from Southeast University, Southeast University, Nanjing, China, in 2007, and 2016, respectively, all in Communication and Information Engineering. He is currently an Associate Professor at the College of Information, Mechanical and Electrical Engineering, Shanghai Normal University, Shanghai 200234, China. 
	
	His research interests include reconfigurable intelligent surface (RIS), cooperative communications, information theory, physical layer security, and machine learning for wireless communications.
	
\end{IEEEbiography}

\vspace{-0.8cm}
\begin{IEEEbiography}[{\includegraphics[width=1in,height=1.25in,clip,keepaspectratio]{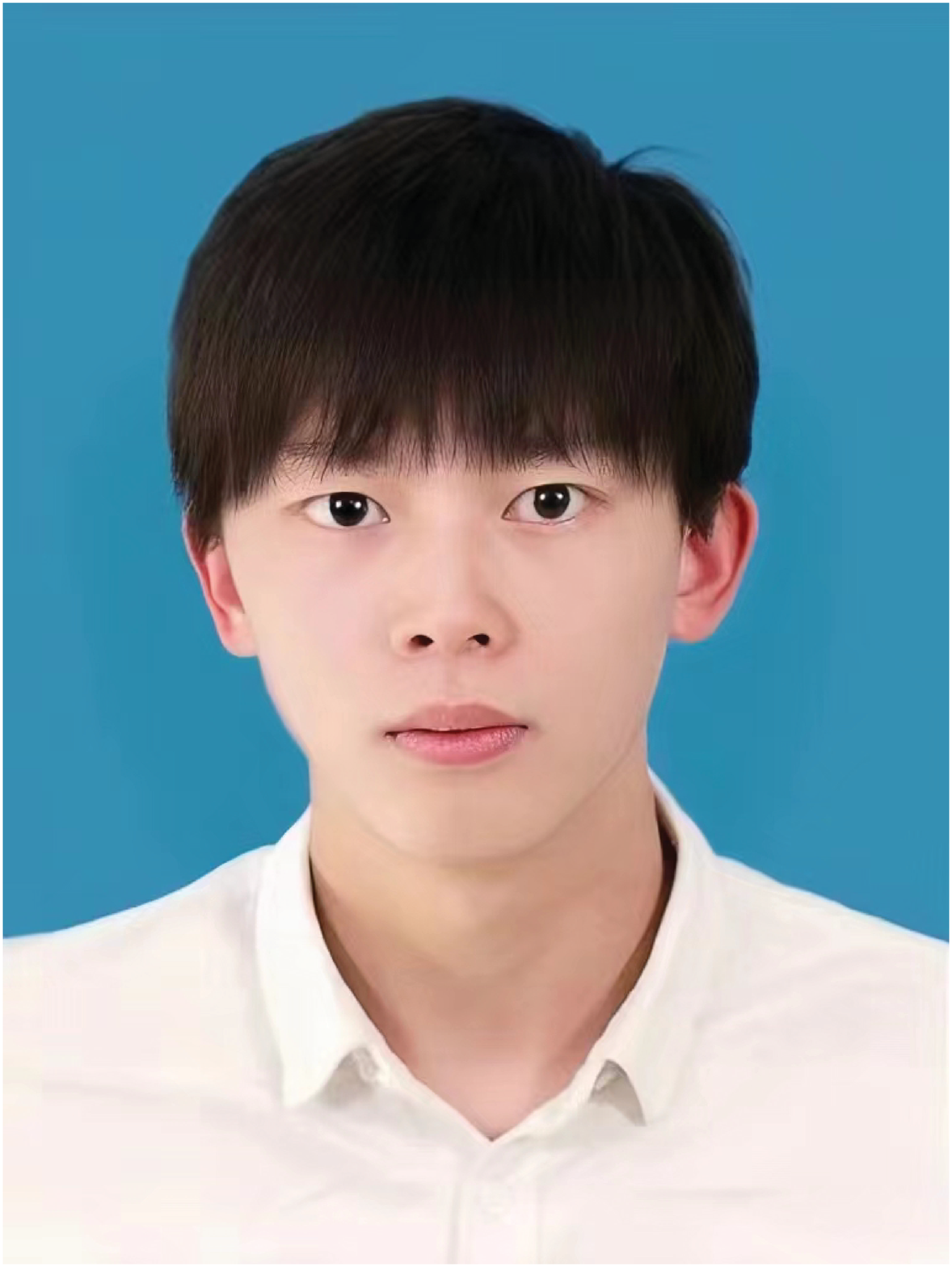}}]{Zhibo Zhang}
	received the B.S. degree from the College of Internet of Things Engineering, Hohai University, Changzhou, China, in 2020. He is currently pursuing the M.S. degree at the College of Information, Mechanical and Electrical Engineering, Shanghai Normal University, Shanghai, China. 
	
	His major research interests lie in the areas of communication and signal processing, including reconfigurable intelligent surface (RIS), physical layer security and machine learning for wireless communications. 
\end{IEEEbiography}

\vspace{-0.5cm}
\begin{IEEEbiography}[{\includegraphics[width=1in,height=1.25in,clip,keepaspectratio]{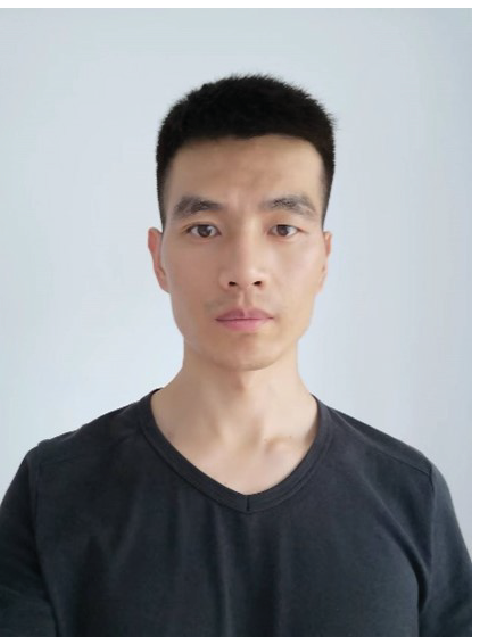}}]{Lei Kong}
	received the Ph.D. degree in the School of Information Science and Engineering, National Mobile Communications Research Laboratory, Southeast University, Nanjing, China in 2016. From 2016 to 2021, he worked as System Engineer at Nokia in Hangzhou, China. He is now a Post Doctor researcher in New H3C Corporation and Southeast University. 
	
	His research interests span on the URLLC, Deterministic Networks, Full Duplex Communication, Network Energy Saving and RIS. 
\end{IEEEbiography}

\begin{IEEEbiography}[{\includegraphics[width=1in,height=1.25in,clip,keepaspectratio]{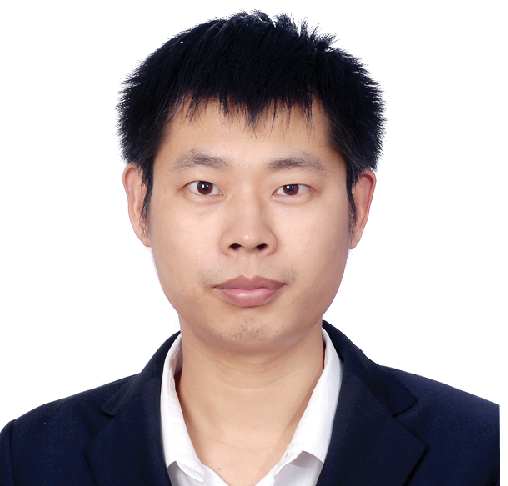}}]{Cunhua Pan}
	(Member, IEEE) received the B.S. and Ph.D. degrees from the School of Information Science and Engineering, Southeast University, Nanjing, China, in 2010 and 2015, respectively. From 2015 to 2016, he was a Research Associate at the University of Kent, U.K. He held a post-doctoral position at Queen Mary University of London, U.K., from 2016 and 2019.From 2019 to 2021, he was a Lecturer in the same university. From 2021, he is a full professor in Southeast University. 
	
	His research interests mainly include reconfigurable intelligent surfaces (RIS), intelligent reflection surface (IRS), ultra-reliable low latency communication (URLLC) , machine learning, UAV, Internet of Things, and mobile edge computing. He has published over 120 IEEE journal papers. He is currently an Editor of IEEE Wireless Communication Letters, IEEE Communications Letters and IEEE ACCESS. He serves as the guest editor for IEEE Journal on Selected Areas in Communications on the special issue on xURLLC in 6G: Next Generation Ultra-Reliable and Low-Latency Communications. He also serves as a leading guest editor of IEEE Journal of Selected Topics in Signal Processing (JSTSP)  Special Issue on Advanced Signal Processing for Reconfigurable Intelligent Surface-aided 6G Networks, leading guest editor of IEEE Vehicular Technology Magazine on the special issue on Backscatter and Reconfigurable Intelligent Surface Empowered Wireless Communications in 6G, leading guest editor of IEEE Open Journal of Vehicular Technology on the special issue of Reconfigurable Intelligent Surface Empowered Wireless Communications in 6G and Beyond, and leading guest editor of IEEE ACCESS Special Issue on Reconfigurable Intelligent Surface Aided Communications for 6G and Beyond. He is Workshop organizer in IEEE ICCC 2021 on the topic of Reconfigurable Intelligent Surfaces for Next Generation Wireless Communications (RIS for 6G Networks), and workshop organizer in IEEE Globecom 2021 on the topic of Reconfigurable Intelligent Surfaces for future wireless communications. He is currently the Workshops and Symposia officer for Reconfigurable Intelligent Surfaces Emerging Technology Initiative. He is workshop chair for IEEE WCNC 2024, and TPC co-chair for IEEE ICCT 2022. He serves as a TPC member for numerous conferences, such as ICC and GLOBECOM, and the Student Travel Grant Chair for ICC 2019. He received the IEEE ComSoc Leonard G. Abraham Prize in 2022.
	
\end{IEEEbiography}

\begin{IEEEbiography}[{\includegraphics[width=1in,height=1.25in,clip,keepaspectratio]{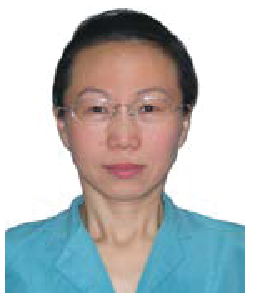}}]{Li Li}
	received the B.S. EE degree from Tsinghua University in 1985, the M.S. EE degree from China Research Institute of Radiowave Propagation in 1988, and Ph.D degree in science from Peking University in 1997. 
	
	Her research interest is currently at the adaptive digital signal processing, spectrum allocation and interference alignment algorithms in heterogeneous cognitive radio network; network modeling of ultra-dense wireless mobile communication.
	
\end{IEEEbiography}

\begin{IEEEbiography}[{\includegraphics[width=1in,height=1.25in,clip,keepaspectratio]{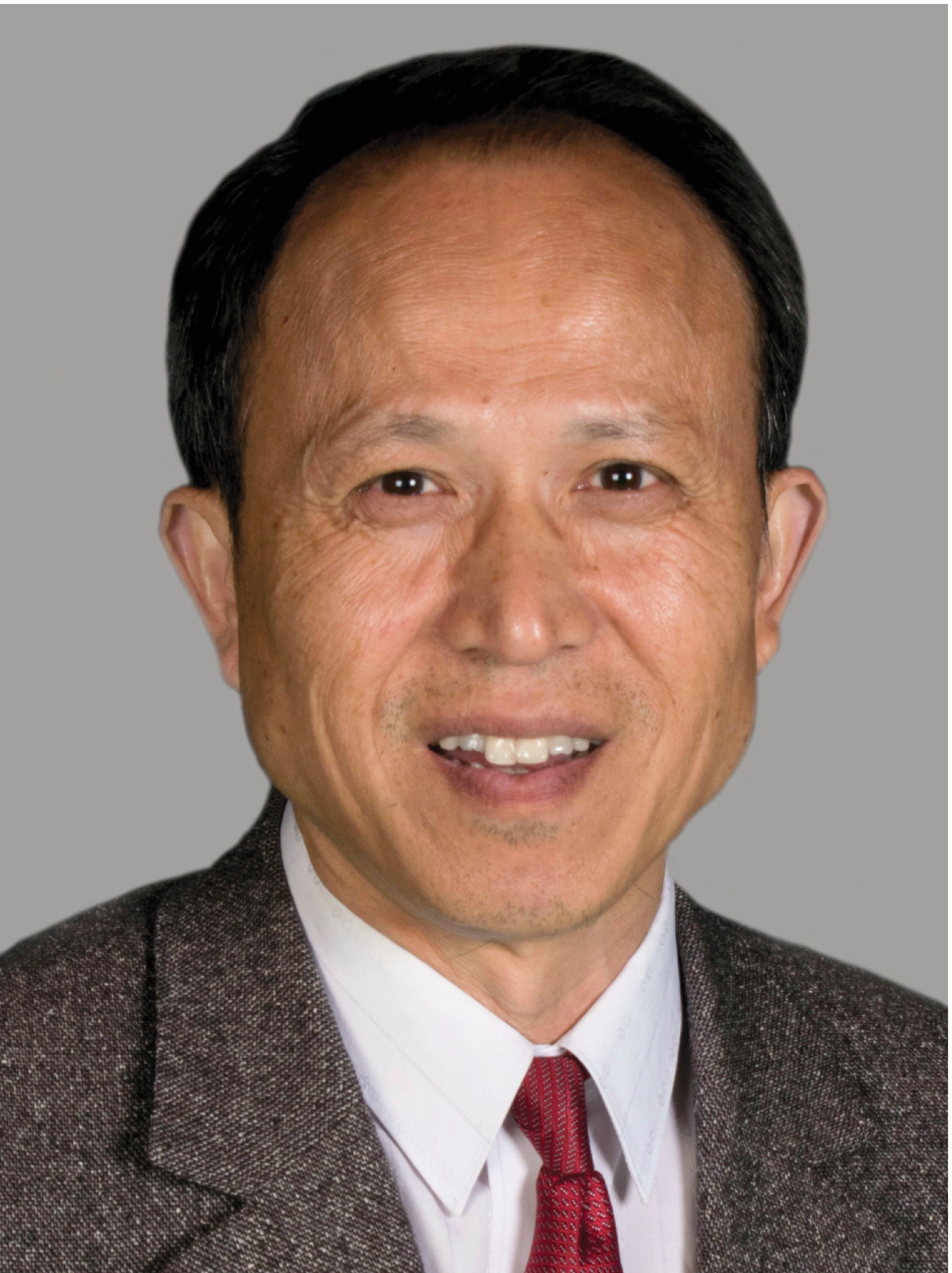}}]{Jiangzhou Wang}
	(Fellow, IEEE) is a Professor at the University of Kent, U.K. His research interest is in the area of mobile communications. He has published over 400 papers and 4 books. He was a recipient of the 2022 IEEE Communications Society Leonard G. Abraham Prize and the 2012 IEEE Globecom Best Paper Award. Professor Wang is a Fellow of the Royal Academy of Engineering, U.K., Fellow of the IEEE, and Fellow of the IET. He was the Technical Program Chair of the 2019 IEEE International Conference on Communications (ICC2019), Shanghai, the Executive Chair of the IEEE ICC2015, London, and the Technical Program Chair of the IEEE WCNC2013. 
\end{IEEEbiography}

\end{document}